\documentclass[fleqn,usenatbib]{mnras}

\usepackage{graphicx}	
\usepackage{amsmath}	
\usepackage{amssymb}	
\usepackage{multicol}        
\usepackage{bm}		
\usepackage{pdflscape}	
\usepackage{xcolor}
\usepackage{listings}
\usepackage{hyperref}

\def\beq{\begin{equation}}
\def\eeq{\end{equation}}
\newcommand{\bea}{\begin{eqnarray}\begin{aligned}}
\newcommand{\eea}{\end{aligned}\end{eqnarray}}

\newcommand{\Gaia}{\textit{Gaia}}
\newcommand{\Galaxia}{\textit{Gaia-Mock}}
\newcommand{\vm}{\textsc{VM2}}
\newcommand{\via}{\textsc{Via Machinae}}
\newcommand{\streamfinder}{\textsc{Streamfinder}}

\usepackage[T1]{fontenc}
\usepackage{ae,aecompl}

\usepackage{newtxtext,newtxmath}

\title[Via Machinae 2.0]{{\textsc{Via Machinae 2.0}: Full-Sky, Model-Agnostic Search for Stellar Streams in \Gaia{} DR2}}
\author[D.~Shih, M.~R.~Buckley, and L.~Necib]{
David Shih,$^{1}$\thanks{E-mail: shih@physics.rutgers.edu}
Matthew R.~Buckley,$^{1}$
and Lina Necib$^{2,3}$
\\
$^{1}$NHETC, Dept. of Physics and Astronomy, Rutgers, Piscataway, NJ 08854, USA\\
$^{2}$Department of Physics and Kavli Institute for Astrophysics and Space Research,
Massachusetts Institute of Technology,\\
77 Massachusetts Ave, Cambridge MA 02139, USA\\
$^{3}$The NSF AI Institute for Artificial Intelligence and Fundamental Interactions,
Massachusetts Institute of Technology,\\
77 Massachusetts Ave, Cambridge MA 02139, USA\\
}

\pubyear{2023}

\begin{document}
\label{firstpage}
\pagerange{\pageref{firstpage}--\pageref{lastpage}}
\maketitle

\begin{abstract}
We present an update to \textsc{Via Machinae}, an automated stellar stream-finding algorithm based on the deep learning anomaly detector ANODE. 
\textsc{Via Machinae} identifies stellar streams within \Gaia{}, using only angular positions, proper motions, and photometry, without reference to a model of the Milky Way potential for orbit integration or stellar distances. 
This new version, \textsc{Via Machinae 2.0}, includes many improvements and refinements to nearly every step of the algorithm, that altogether result in more robust and visually distinct stream candidates than our original formulation. 
In this work, we also provide a quantitative estimate of the false positive rate of \textsc{Via Machinae 2.0} by applying it to a simulated \Gaia{}-mock catalog based on \textit{Galaxia}, a smooth model of the Milky Way that does not contain substructure or stellar streams. 
Finally, we perform the first full-sky search for stellar streams with \textsc{Via Machinae 2.0}, identifying 102 streams at high significance within the \Gaia{} Data Release 2, of which only 10 have been previously identified. 
While follow-up observations for further confirmation are required, taking into account the false positive rate presented in this work, we expect approximately 90 of these stream candidates to correspond to real stellar structures.
\end{abstract}




\defcitealias{2018ApJ...863L..20P}{PWB18}
\defcitealias{Nachman:2020lpy}{NS20}
\defcitealias{full_sky}{Paper II}

\section{Introduction}
\label{sec:introduction}

As dwarf galaxies and globular clusters are tidally stripped by the Milky Way's gravitational field, they form extended stellar structures called streams. The composition \citep[see e.g.][]{2022ApJ...930..103Y,2022ApJ...928...30L}, orbits \citep[see e.g.][]{2021ApJ...923..149S,2022ApJ...928...30L,2022arXiv220802255S}, and internal structure \citep[see e.g.][]{2012ApJ...760...75C,2019ApJ...880...38B}  of these streams are valuable probes into the merger history \citep{1998ApJ...495..297J,1999MNRAS.307..495H,2006ApJ...642L.137B,2018ApJ...862..114S,2018MNRAS.478..611B,2021arXiv210409523M,2022ApJ...926..107M}, structure \citep{1999ApJ...512L.109J,2001ApJ...551..294I,2010ApJ...712..260K,2010DDA....41.0501N,2011MNRAS.417..198V,2012JCAP...08..027P,2013MNRAS.433.1813S,2015ApJ...803...80K,2019MNRAS.486.2995M,2019ApJ...883...27N,2020arXiv200700356R,2022MNRAS.515.3818T}, and substructure of the Galaxy \citep{2012ApJ...760...75C,2016MNRAS.457.3817S,2017MNRAS.470...60E,2019ApJ...880...38B,2019MNRAS.484.2009B,2020ApJ...892L..37B}. As the number of wide field and all-sky surveys has grown, the number of known stellar streams has likewise increased \citep[see][for a summary]{2022arXiv220410326M}. Most recently and notably, the \Gaia{} Space Telescope has opened a new frontier in the discovery and study of stellar streams, providing both all-sky coverage and proper motion measurements for 1.5 billion stars \citep{2016A&A...595A...1G,2016A&A...595A...2G,2018A&A...616A...1G,2021A&A...649A...1G,2022arXiv220800211G}. 

A number of automated techniques have been developed to search for stellar streams within the \Gaia{} data \citep[see e.g.][]{2018MNRAS.477.4063M,2018MNRAS.478.3862M,2018ApJ...863...26Y,2019A&A...621L...3M,2020MNRAS.492.1370B,2019A&A...622L..13M,2020arXiv201205245I}.
Generally speaking, existing approaches require some model-dependent assumptions about the dynamics and astrophysics of the Milky Way.
For example, the \textsc{Streamfinder} algorithm \citep{2018MNRAS.477.4063M} -- by far the most comprehensive automated approach to stream finding to date -- identifies stellar streams by searching the Milky Way for stars that occupy the same position/velocity ``hypertube'' in their projected orbits through the Galaxy; this requires an assumption for the Galactic potential, something that is still uncertain and a topic of current research \citep[see e.g.][]{2022arXiv221104495K,2023MNRAS.518..774L}. 

\begin{figure*}
\begin{centering}
\includegraphics[trim={0 6.7cm 0 6cm},clip,width=0.95\textwidth]{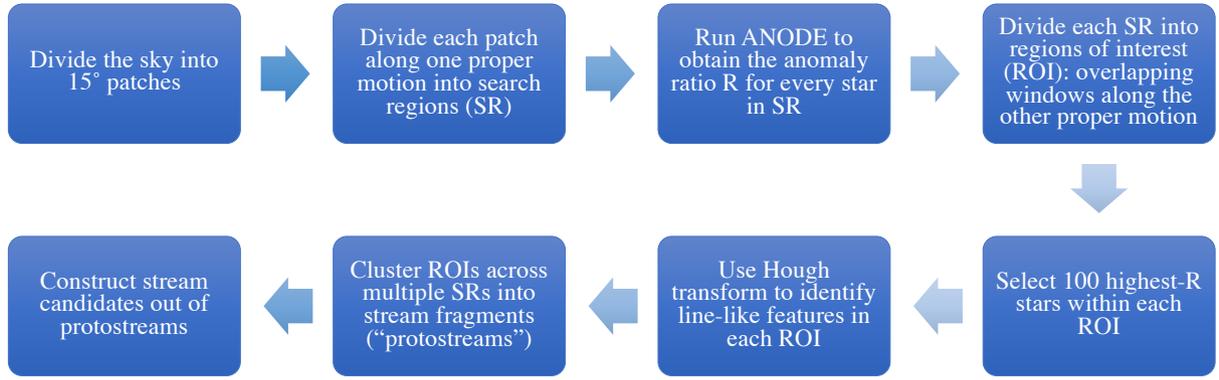}
\caption{Flowchart of \vm{}. In this updated version, numerous steps, including the line-finding algorithm, and the stages whereby the ROIs are clustered into full stream candidates, have been improved in order to produce higher quality stream candidates, as well as to minimize the false positivity rate. Also, in \vm{} we have run ANODE on both orthogonal proper motions instead of just a single one; this increases the chances and robustness of stream detections. See Sections~\ref{sec:inputs} and \ref{sec:vm2} for details.}
\label{fig:flowchart}
\end{centering}
\end{figure*}

In a prior paper \citep[][hereafter referred to as Paper I]{2022MNRAS.509.5992S}, we developed \textsc{Via Machinae}: a new, fully-automated, machine-learning-assisted algorithm to discover stellar streams within the \Gaia{} data. Unlike other stream finding algorithms, \textsc{Via Machinae} is {\it model-agnostic}, in that it does not assume anything about the form the Galactic potential, orbits, or isochrones.

\textsc{Via Machinae} is based on the ANODE algorithm \citep{Nachman:2020lpy}, a simulation-independent, data-driven, unsupervised machine learning method for anomaly detection originally developed for the Large Hadron Collider. ANODE uses normalizing flows (for reviews and original references, see \cite{2019arXiv190809257K,2019arXiv191202762P}) to estimate the probability density of data in a ``search region," as well as a background probability density from sidebands.  (See \cite{Nachman:2020lpy} and Paper I for more details.) Applied to the angular positions, proper motions, color and magnitude of the stars in the \Gaia{} catalog, \textsc{Via Machinae} first uses ANODE to identify stars that are anomalous (overdense) with respect to the background. Subsequent steps of \textsc{Via Machinae} aim to specialize to overdensities that are broadly consistent with stellar streams: localized in both proper motion coordinates, with locally line-like spatial extent on the sky, and color and magnitudes consistent with old, metal-poor stars, but not necessarily following a specific isochrone.

In Paper I, we developed the steps of \textsc{Via Machinae}, and demonstrated their efficacy in rediscovering GD-1  \citep{2006ApJ...643L..17G}, a particularly extended, narrow, and dense stellar stream. 
While the basic outline of \textsc{Via Machinae} has not changed from Paper I, in this work we present many refinements and improvements to the algorithm in order to improve the quality of the discovered stream candidates and its robustness against false positives. 
 We will refer to the new-and-improved version of \textsc{Via Machinae} as \vm{} throughout this work. 

Figure~\ref{fig:flowchart} shows the schematic of the steps of the \vm{} algorithm, which are:
\begin{enumerate}
\item We divide the sky into overlapping patches, $15^\circ$ in radius. Patches too close to the Galactic disk are discarded, as are patches containing the Large Magellanic Cloud or significant dust occlusions. Each patch has its own centered angular coordinates $\lambda$ and $\phi$, obtained from rotating the ICRS right ascension ($\alpha$) and declination ($\delta$) coordinates. The rest of the analysis is done in these patch-centered coordinates.

\item The stars in each patch are then divided into overlapping {\it search regions} according to one of their two proper motion coordinates $\mu_\lambda$ and $\mu_{\phi^*}$. These regions have a set width of 6~mas/yr, and the centers of the overlapping regions are separated by 1~mas/yr. The complement of each search region defines its paired {\it control region}. In \vm{}, we perform separate scans with search regions defined in $\mu_\lambda$ and again in $\mu_{\phi^*}$. Only the first proper motion coordinate was used in the analysis of Paper I.

\item The ANODE algorithm is then run on each search region, using the associated control region to determine a background probability density. The result is an anomaly score for each star in the search region:
\begin{equation} \label{eq:R}
R(x) = \frac{p_{\rm SR}(x)}{p_{\rm bg}(x)},
\end{equation}
where $p_{\rm{SR}}(x)$ is the probability density of the star in the signal region, and $p_{\rm{bg}}(x)$ is the background probability density of the star, obtained by interpolating from the control region.
Here $x$ are all the features of the star used in training ANODE, which for \textsc{Via Machinae} consists of $x=(\lambda,\phi,\mu',g,b-r)$, with $\mu'$ being the orthogonal proper motion coordinate that was not used to define the search region, $g$ the \Gaia{} magnitude, and $b-r$ the \Gaia{} color.

We note that each star will typically appear in multiple search regions and patches, since these are highly overlapping. Therefore, each star will typically have multiple $R(x)$ values. As the anomaly score depends on the exact selection of stars within the search and control regions (as well as stochastic dependence on the training process), each $R(x)$ constitutes approximately-independent tests for the anomalous properties of the star. Below, we will describe how these multiple quasi-independent $R(x)$ values are combined to build more robust, higher-confidence stream candidate detections. 

\item We apply a set of \textit{fiducial selection cuts}, selecting stars whose angular coordinates are within the inner $10^\circ$ of the patch, $g <20.2$, and $0.5<b-r<1$. To these cuts -- which were implemented in Paper I -- we add a new cut, designed to remove disk stars, which act as foreground in the search of stellar streams located in the stellar halo. For stars with well-measured parallaxes, we require their Galactocentric altitude $|z|$ to be greater than 2~kpc at 95\% CL (the full details of this new criteria are described below). These fiducial cuts are imposed {\it after} the ANODE training, as hard boundaries within the data set would result in errors in the density estimation near the edges.

\item We further subdivide each search region into overlapping regions of interest (ROIs) by selecting windows of width 6~mas/yr in the orthogonal proper motion coordinate that was not used to define the search region (e.g., for search regions defined in $\mu_\lambda$, the ROIs are constructed by slices in $\mu_{\phi^*}$).\footnote{Note this results in two sets of ROIs for each patch that have the same boundaries in $\mu_\lambda$ and $\mu_{\phi^*}$. However, as the training of ANODE occurs before the division into ROIs, the $R$ values of these two sets are distinct.} Within each ROI, we select the 100 stars with the highest $R(x)$ values. 

\item Using a Hough transform \citep{Hough:1959qva,10.1145/361237.361242}, we perform an automated line-finding on the angular positions of these 100 stars for each ROI. This line-finding process returns, in addition to the location and orientation of the best-fit line, a statistical significance for the line. Compared to Paper I, this step has been modified to better estimate the background and statistical significance of the most-probable line of stars within any ROI.

\item We expect a real stream to appear as an anomaly within multiple overlapping ROIs in a given patch. We therefore cluster together the lines from ROIs whose line location, orientation, and proper motions are concordant. We refer to these combined ROIs as ``protoclusters.'' The combination process also results in a single protocluster significance parameter obtained from re-running the line-finding on the concatenation of all the ROIs in the protocluster. Compared to Paper I, the algorithm for this line combination step has been greatly improved in \vm{}, with more flexible and accurate measures of concordance between ROIs, resulting in fewer spurious combinations. We also now combine lines from the ROIs trained on $\mu_\lambda$ and $\mu_{\phi^*}$ within the same patch at this step, which was not possible in Paper I given that the analysis there was conducted in a single proper motion parameter.

\item Finally, we construct stream candidates by clustering the protoclusters from different patches into a single set of stars on the sky, again requiring a consistent set of proper motion values and line directions across the stream candidate. From the protocluster significances, a single stream significance parameter is calculated, which can be used to reject false positives. Again, as compared to Paper I, the algorithm used in this step is improved to more accurately combine the fragments of a stream located in different patches, taking into account the fact that the proper motion and apparent direction of the stream on the sky can change along its path.
\end{enumerate}

In addition to the improvements to \textsc{Via Machinae} described above, this paper contains a quantitative estimate the false positive rate (fpr) of \textsc{Via Machinae} for stream finding,
using the set of \Gaia-like observations described in \cite{2018PASP..130g4101R}. These synthetic stars were drawn from a completely smooth model of the Milky Way with no substructure, thus any stream candidates our algorithm detects in this data set are spurious.

Finally, in this paper we present the first results from an all-sky scan of the \Gaia{} Data Release 2 (DR2) dataset using \textsc{Via Machinae}. Altogether, we find 102 stream candidates, and with the fpr derived for this working point, we may expect $\sim 90$ of the stream candidates to be real stellar streams. \vm{} is able to rediscover six previously known streams, as well as fragments of Sagittarius. The rest of the stream candidates found by \vm{} do not correspond to known streams. We briefly describe 15 of the most-significant new stream candidates identified using \vm{}, saving a more detailed study for an upcoming follow-up work.

The remainder of the paper is organized as follows. In Section~\ref{sec:inputs}, we describe the \Gaia{} data used in this analysis, as well as our methods to detect and remove regions which are too contaminated with dust, globular clusters, or thick disk stars for our stream-finding algorithm to return sensible results. Section~\ref{sec:vm2} explains in detail the changes and improvements to the \textsc{Via Machinae} algorithm from Version~1 to the current Version~2. In Section~\ref{sec:galaxia}, we estimate the false-positive rate of \textsc{Via Machinae} using the \Gaia{} mock-catalog. Lastly,  we apply \vm{} to the entire \Gaia{} DR2 sky and describe our results in Section~\ref{sec:rediscovery}. 

We note that the technical details of our data processing in Section~\ref{sec:inputs} and of our algorithm in Section~\ref{sec:vm2} are rather dense and may not be of interest to all readers. The high-level description of the \vm{} algorithm provided here in the Introduction is intended to be sufficient to understand the algorithm for most readers, and provide the necessary context for the fpr calculations in Section~\ref{sec:galaxia} and the stellar streams identified in Section~\ref{sec:rediscovery}. 

\section{Data set}
\label{sec:inputs}

\subsection{Overview}
\label{sec:dataoverview}

\begin{figure*}
\begin{centering}
\includegraphics[trim={0 2.5cm 3cm 2cm},clip,width=0.75\textwidth]{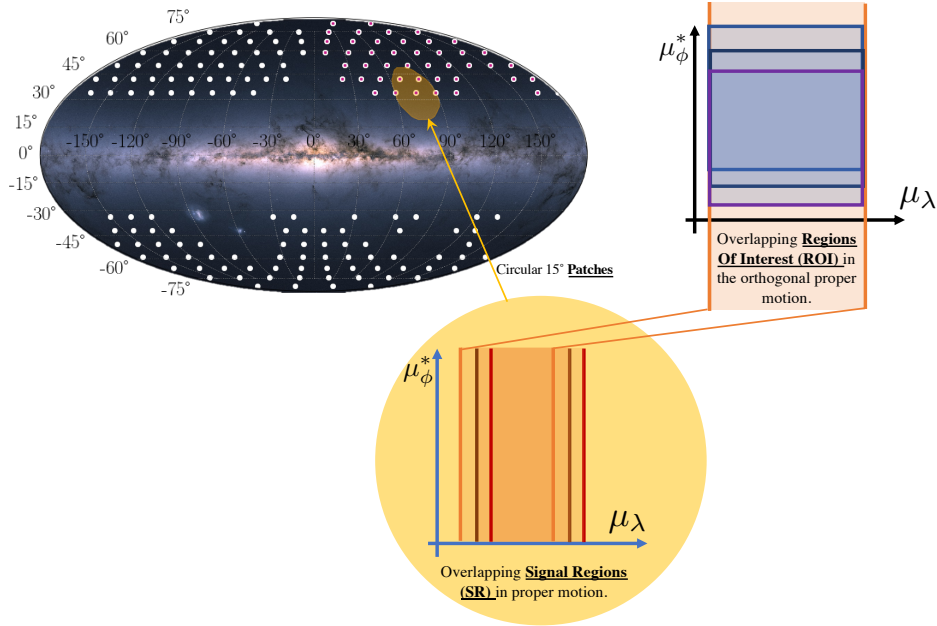}
\caption{Definition of patches, signal regions (SRs), and regions of interest (ROIs). See Section~\ref{sec:dataoverview} for details. The dots on the sky map indicate the centers of the patches considered in the analysis. The analysis was run on the full sky with \Gaia{} (see Sec~\ref{sec:rediscovery}), and on the quarter of sky shown with red dots with \Galaxia{} (see Sec~\ref{sec:galaxia}).  }
\label{fig:regions}
\end{centering}
\end{figure*}

In Paper I, we demonstrated the \textsc{Via Machinae} method using a data set derived from \Gaia{} DR2. While the current work was already underway,  \Gaia{} early Data Release 3 (eDR3) and Data Release 3 (DR3) was published. As we do not use parallax, radial velocity, or spectroscopic information from individual stars in the machine learning stage of our algorithm, the primary advantage of (e)DR3 data over DR2 for our purposes is the greatly improved and reduced measurement errors, which likely reduces false positive stream detections and improves the purity of our stream candidates with respect to contamination by non-stream stars. However, given that significant computational resources are required for the machine learning algorithm, we continue to present results from DR2 in this paper. In addition, the results of a \via{} scan on DR3 would not completely supersede our DR2 results; rather, given the stochastic nature of the training of the machine learning algorithm, evidence for streams provided by these scans could be considered quasi-independent: we expect that real streams would robustly show up in both DR2 and DR3 scans, while spurious false positives in DR2 would be less robust and might not be confirmed by DR3. An analysis of the DR3 data set is forthcoming in a future work.

The data we use in the \via{} method (both here and in Paper I) is organized into three nested levels: patches, signal regions (SRs), and regions-of-interest (ROIs). We now review their definitions in more detail (see Figure~\ref{fig:regions} for a graphical illustration):

\begin{itemize}
\item {\it Patches:} Starting with \Gaia{} DR2, we select stars with measured parallax less than 1~mas. We do not correct for the zero-point parallax offset as we do not use the detailed parallax information, and therefore we do not expect this to have any significant effect on our analysis. We construct overlapping circular patches on the sky of radius $15^\circ$, with patch centers in Galactic coordinates selected using \textsc{HealPy} \citep{2005ApJ...622..759G,Zonca2019} (with \texttt{nside=5}). Patches near the disk contain a large number of stars, and data can be highly affected by the presence of dust. The former issue results in prohibitively long training at the ANODE step, and the latter causes spurious anomalous features. For these reasons, we remove all patches with centers that have $|b| <30^\circ$. We further exclude those that overlap with the Large or Small Magellanic Clouds, as well as those patches away from the disk that contain significant dust features (these are primarily located near the Galactic bulge). We are left with a total of 163 patches overall, which form the basis for our all-sky scan. (This should be contrasted with Paper I, which focused on just 22 patches containing GD-1.)

We select the angular position, proper motion, photometric magnitude $g$, and photometric $b-r$ color as features of each star, to be used in the ANODE training. In each patch, we recenter the ICRS coordinate system to a set of local angular coordinates $\phi$ and $\lambda$ using the built-in centering function of the \texttt{SkyOffsetFrame} object in the \textsc{Astropy} package \citep{astropy:2013,astropy:2018,astropy:2022}. The associated proper motions are likewise rotated from the ICRS coordinate system to the patch-centered coordinates: $\mu_\lambda$ and $\mu_{\phi^*}~\equiv~\mu_\phi~\cos\lambda$. 
These steps are all the same as in Paper I.

\item {\it Signal regions (SRs):} Within each patch, we divide stars into overlapping search SRs in proper motion space. In Paper I, we used only the $\mu_\lambda$ coordinate to define the SRs; here, we construct two sets: first using $\mu_\lambda$, and then a separate set using $\mu_{\phi^*}$. In each case, the SRs are defined as those stars within the given proper motion's range of 6~mas/yr, with the center of each SR separated by 1~mas/yr from its neighbors (see Fig.~\ref{fig:regions}).:
\begin{equation}
\left[ \mu_{\rm min}, \mu_{\rm max}\right] = \ldots,[-10,-4],[-9,-3],\ldots,[3,9],[4,10],\ldots
\end{equation} 
SRs with fewer than $2\times 10^4$ stars or more than $10^6$ stars are excluded from further analysis. At the low end, this number of stars is not sufficient for accurate ANODE training; at the high end the training times become prohibitively long. Each SR implicitly defines its control region (CR), the compliment of stars outside the selected range of proper motions. In summary, there are 5,152~SRs in $\mu_\lambda$ and 5,465~SRs in $\mu_{\phi^*}$. That corresponds to an average of 33~SRs in each patch and each proper motion coordinate.

The SRs and their CR counterparts are fed into the ANODE training step, described in detail in Paper I and briefly reviewed in section~\ref{sec:ANODE}. ANODE produces an overdensity score $R(x)$ for every star in the SR; by cutting on this score, we can discover stellar streams hidden in the data. This is the core of the \via{} method.  

\item {\it Regions of Interest (ROIs):} After running the ANODE method on the SRs, and before running the subsequent steps of \via, we further divide up the SRs into ROIs by cutting on the other, orthogonal proper motion coordinate. In other words, if the SR is defined using $\mu_\lambda~(\mu_{\phi^*})\in[n,n+6]$~mas/yr then the ROI is defined by additionally restricting $\mu_{\phi^*}~(\mu_\lambda)\in[m,m+6]$~mas/yr. Like SRs, the cuts on the orthogonal proper motion coordinate are also defined with width 6~mas/yr and stepsize 1~mas/yr, so the ROIs are overlapping square regions in proper motion space (see Fig.~\ref{fig:regions}). The purpose of further dividing up the data into ROIs is that streams are supposed to be localized in both proper motion coordinates, so by restricting to the stars in an ROI, we hope to enhance the signal to noise ratio of any stream that might be present. All in all, there are
126,727~ROIs corresponding to the $\mu_\lambda$ SRs and 129,830~ROIs corresponding to the $\mu_{\phi^*}$ SRs. 

\end{itemize}

Finally, after dividing the data into ROIs, two more selection criteria are required before running the subsequent steps of \vm{}: identifying and excluding ROIs that contain globular clusters or dwarf galaxies, and defining a fiducial region in which we perform the rest of the analysis; this region is chosen to maximize the signal-to-noise of stream candidates. Since both of these steps are considerably changed from Paper I, we will describe them in more detail in the following two subsections.

\subsection{New Globular Cluster/Dwarf Galaxy finding algorithm} \label{sec:GCfinding}

Like stellar streams, globular clusters (GCs) and dwarf galaxies (DGs) are collections of stars in compact regions of kinematic phase space, and are therefore anomalous compared to the background. 
The difference lies in the clustering of these objects in different spaces: Streams are extended in one angular direction and localized in the orthogonal direction as well as the proper motions. GCs and DGs, however, are compact in all angular and proper motion coordinates. Due to the large number of stars within a GC/DG and their anomalous properties as identified by ANODE, we find that the presence of a GC or DG within a search region can cause complete failure of our later steps in identifying stream candidates. 
This issue was identified in Paper I, but our initial algorithm to remove GCs and DGs was not sufficiently robust for the full-sky analysis. We provide the revamped analysis here.

\begin{figure}
\begin{centering}
\includegraphics[trim={0cm 1.5cm 1cm 2cm},clip,width=0.48\textwidth]{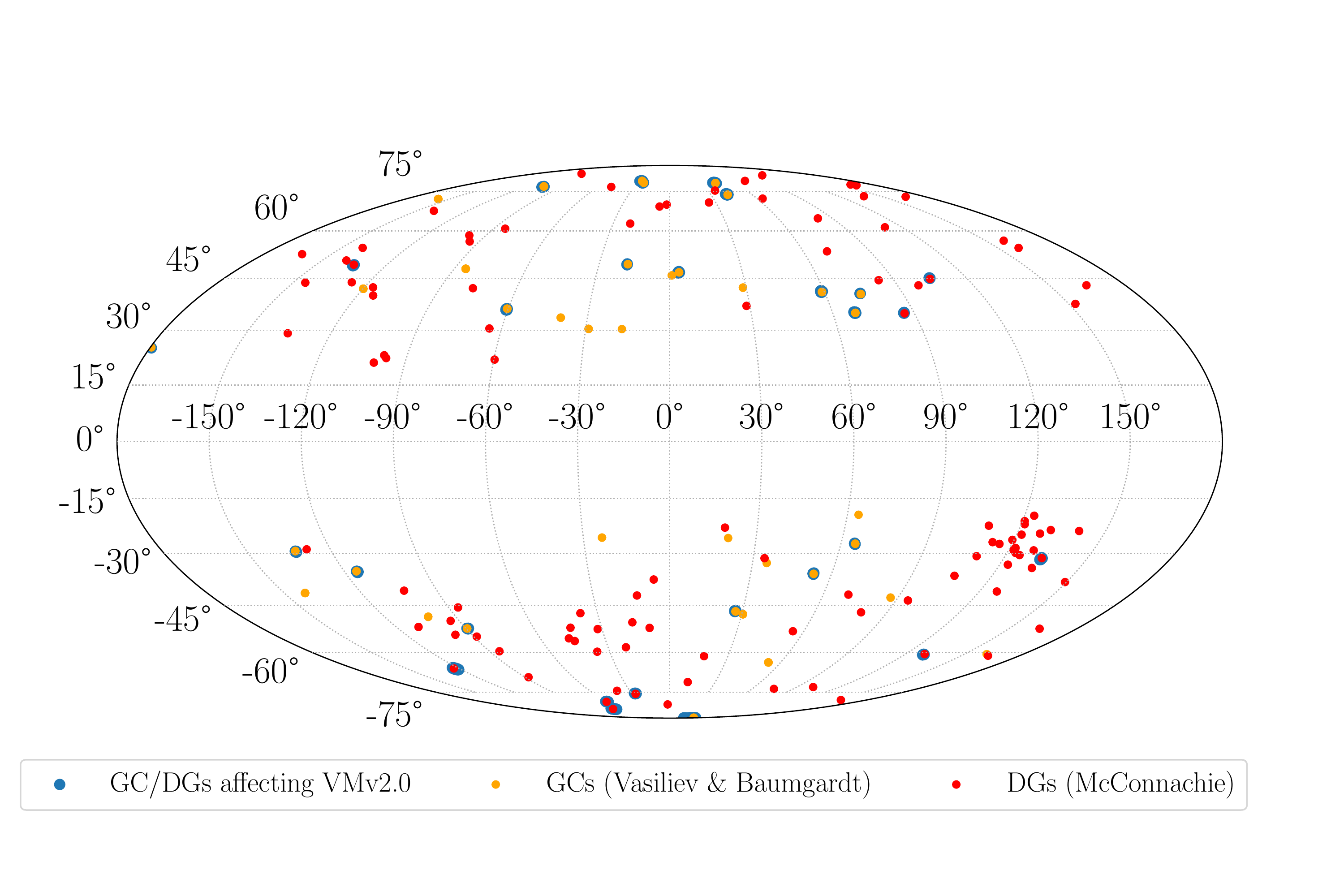}
\caption{The GC/Dwarf galaxy candidates identified and removed in our analysis (blue), overlaid with the known GCs from \protect\cite{2021MNRAS.505.5978V} and DGs from \protect\cite{2012AJ....144....4M}. We only show the GCs and DGs contained in the analyzed region of this work (163 patches of the \Gaia{} DR2 scan, as defined in Sec.~\ref{sec:dataoverview}). This figure confirms that every GC/DG identified by our algorithm as affecting \vm{} is indeed a known object. The converse is of course not true -- the vast majority of GCs and DGs are too dim and/or far away to even show up in the \Gaia{} DR2 catalog.}
\label{fig:GCcandidates}
\end{centering}
\end{figure}

To locate the GC/DG overdensities, we make a two-dimensional histogram of all the stars within a SR in the angular coordinates $\phi$ and $\lambda$, using $120\times 120$ bins (enumerated by $i$ and $j$ in the $\phi$ and $\lambda$ coordinates respectively) to cover the full $15^\circ$ patch. The goal is to identify significant overdensities in the number of stars within each pixel, $N_{ij}$, as compared to neighboring pixels. Given that this set up is similar to identifying line-like structures in Hough space, we develop a common algorithm that can be adapted for both problems. This algorithm is described in detail in Appendix~\ref{app:annulus}; briefly, the overdensity in each pixel is quantified as the difference between $N_{ij}$ and the average counts in an annulus centered on the $i,j$ pixel, normalized to the statistical and systematic errors. For the GC/DG-finding, we use an annulus with a width and height of 11 pixels, minus the inner $3\times 3$ pixels. The normalized significance assigned to each pixel by this procedure is labeled $S_{ij}$.

We find that pixels with $S_{ij}\lesssim9$ are indistinguishable from noise by eye, whereas pixels with $S_{ij}\gtrsim50$ are clearly recognizable as localized objects. In between, we find that not every bright pixel with $9<S_{ij}<50$ causes ANODE to fail. We therefore construct a two-step cut as follows. Given a GC/DG candidate pixel with $S_{ij}>9$, we consider every ROI $r$ in the SR. Let $N_{r,ij}$ be the number of the 100 highest-$R$ stars in the ROI localized within a $0.5^\circ$ circle around the GC/DG candidate.
We then exclude SRs containing pixels that satisfy
\beq
S_{ij}>50 \qquad {\rm OR} \qquad (S_{ij}>9 \quad {\rm AND}\quad \max_r N_{r,ij}>10).
\eeq
In other words, we exclude SRs containing the highest significance pixels, as well as the pixels with somewhat lower significance but that contain a high  number of anomalous stars.
In all, 1,174 (1,098) SRs are excluded from the analysis, leaving 3,978 (4,367) SRs corresponding to the $\mu_\lambda$ ($\mu_{\phi^*}$) directions.

A map of all the GC/DG candidates identified and removed by our analysis is shown in Figure~\ref{fig:GCcandidates}, overlaid with all known GCs \citep{2021MNRAS.505.5978V} and DGs \citep{2012AJ....144....4M} that are contained within the 163 patches in our \Gaia{} DR2 scan. We see that every object identified and removed by our method corresponds to a known GC or DG, but there are many more existing GCs and DGs not removed by our analysis. This is due in part to the fact that our tests have shown that not every bright localized GC/DG-like object caused ANODE to fail, and we opted therefore not to remove these regions from our analysis. Furthermore, many GCs and DGs are simply too faint or distant to be detectable in the Gaia DR2 data set.

\subsection{New fiducial region: excluding thick disk stars}
\label{sec:datasetfid}

In Paper I, we imposed several fiducial cuts post-ANODE training based on the quality of the density estimation (avoiding sharp edges in probability distributions),  namely $g<20.2$ and $r<10^\circ$. We also removed stars with both proper motions too close to zero ($|\mu_\lambda|<2$~mas/yr and $|\mu_{\phi^*}|<2$~mas/yr), to remove a population of very distant stars whose presence would overwhelm the ANODE anomaly finder. Finally, we
required $(b-r)\in [0.5,1]$ after ANODE training, in order to target cold stellar streams containing old, metal-poor stars.

In this work, we do not target close streams within the disk \citep[for stream searches that do target this region of the Milky Way, see for example][]{1999MNRAS.307..495H,2018MNRAS.475.1537M,2018MNRAS.478.5449M,2020NatAs...4.1078N}. To avoid foreground contamination, particularly that of the Galactic disk, we therefore exclude nearby stars. Significant contamination from bright foreground stars can indeed by seen in the initial formulation of the \textsc{Via Machinae} algorithm: Figure~13 (lower right panel) of Paper I shows a population of bright ($G\lesssim 16$) stars superimposed on our detection of GD-1.
This contamination is not unique to the GD-1 stream -- disk stars are spatially and kinematically coherent, which makes it likely that ANODE will identify them as anomalous. 

\begin{figure}
\begin{centering}
\includegraphics[width=0.9\columnwidth]{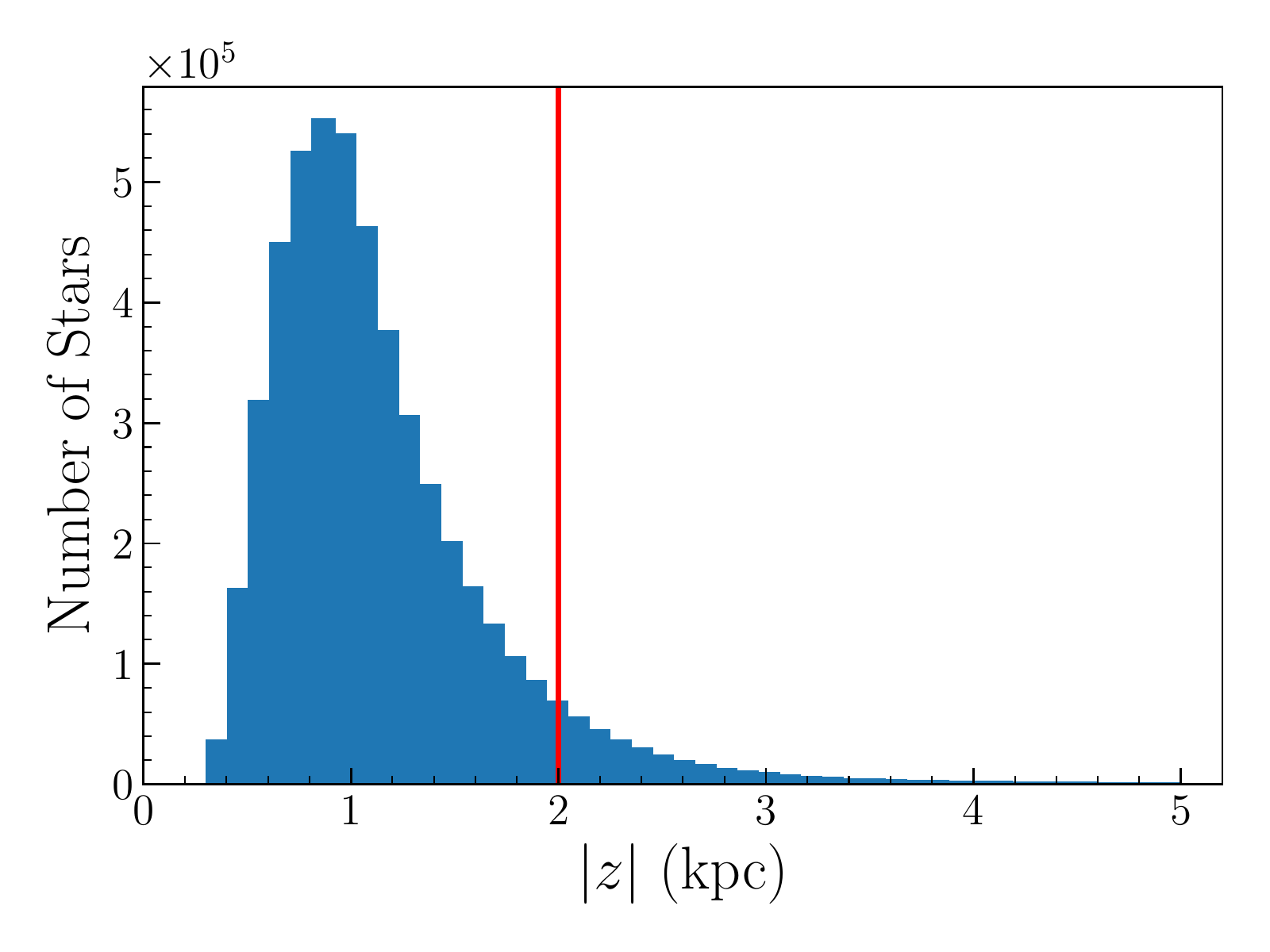}
\caption{Absolute value of Galactocentric $z$, for all stars with well-measured parallax (Eq.~\ref{eq:parallax_cut}) in our data set.}
\label{fig:galz}
\end{centering}
\end{figure}

Fortunately, these stars can be removed using the following new fiducial cut that we impose. This cut focuses only on stars that have ``well-measured parallaxes" $\varpi$, which we define as
\beq \label{eq:parallax_cut}
\varpi > 0\quad {\rm and} \quad {\sigma_\varpi\over\varpi} <0.2,
\eeq
where $\sigma_\varpi$ is the measured parallax error. Approximately 16\% of the stars across our 163 patches of \Gaia{} DR2 satisfy these requirements. 
For these stars, we have enough information to transform their coordinates to cartesian Galactic coordinates. In Figure~\ref{fig:galz}, we plot the vertical distance $|z|$ (altitude relative to the Galactic midplane) of these stars. As expected, the vast majority of these stars have low values of $z$, implying that they are predominantly disk stars.\footnote{The absence of stars with $|z|$ close to zero is due to our initial cuts on parallax $\varpi<1$~mas and Galactic latitude $|b|>30^\circ$.} 
 To remove these stars from our stream search, we impose the new fiducial cut
\beq\label{eq:zcut}
|z_{95}|>2~{\rm kpc}
\eeq 
on all stars with measured parallaxes. Here, $|z_{95}|$ is the $z$ coordinate obtained by transforming the 95\% upper limit on the parallax, $\varpi+2\sigma_\varpi$, to Cartesian coordinates.\footnote{Using the upper limit instead of the mean value of the $z$ coordinate is a tighter fiducial cut as $|z_{95}| < |z|$.}  The specific value of 2~kpc was chosen to correspond to approximately twice the scale height of the Milky Way thick disk \citep{2017ApJ...850...25L}. This selection is imposed after running ANODE, along with our other existing fiducial cuts described above.

\begin{figure}
\begin{centering}
\includegraphics[width=0.95\columnwidth]{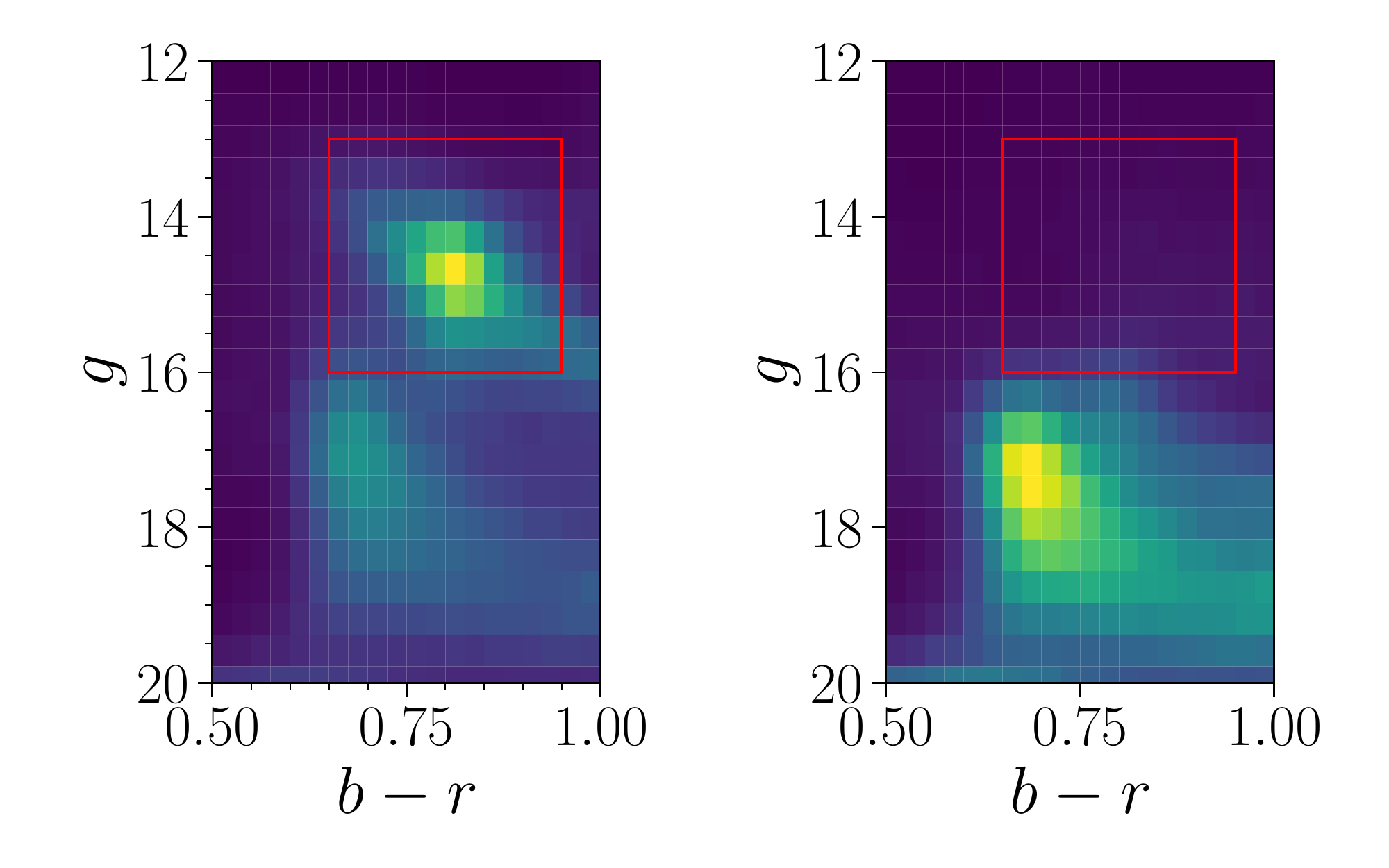}
\caption{Left: Observed magnitude $g$ versus $b-r$ color for the collection of 100 highest-$R$ stars from all ROIs across the entire \Gaia{} data set. Red rectangle indicates bright stars that are contamination from the Galactic disk. Right: The same but after the selection on the vertical distance $z$ given by Eq.~\eqref{eq:zcut}.  }
\label{fig:photometryROI}
\end{centering}
\end{figure}

In Figure~\ref{fig:photometryROI}, we illustrate the effectiveness of this cut with a two-dimensional histogram of the color and magnitude of the 100 highest-$R$ stars in every ROI across our entire \Gaia{} data set. The left panel shows all the stars, while the right panel only shows the stars remaining after the selection of Eq.~\eqref{eq:zcut}. The red rectangle is added to visually emphasize the location of the bright disk stars removed by this fiducial cut on $|z_{95}|$.

\section{Via Machinae 2.0} \label{sec:vm2}

\subsection{Initial steps}
\label{sec:ANODE}

We now describe the \textsc{Via Machinae} algorithm in detail. An overview has been provided in the Introduction (Sec.~\ref{sec:introduction}), and we recapitulate some of that information here. We begin by describing the initial steps of \vm{}. These steps are mostly unchanged from Paper I where we refer the reader for more details.

For each SR in each circular 15$^\circ$ patch defined in Section~\ref{sec:dataoverview},
we run the less-than-supervised anomaly detection algorithm known as ANODE \citep{Nachman:2020lpy}. The input features to ANODE are $\vec{x}=(\lambda,\phi,\mu',g,b-r)$, where $\mu'$ is the orthogonal proper motion to the one used to define the SR (for more details, see Section~\ref{sec:dataoverview}). ANODE is based on a technique for density estimation known as normalizing flows (for recent reviews, see \cite{9089305, papamakarios2021normalizing}), in particular the architecture of Masked Autoregressive Flows (MAF) \citep{NIPS2017_6c1da886}. For full details of the MAF architecture, see Paper I.

By training ANODE on the stars in the SR and then interpolating the sideband density estimate from the control region into the SR, we obtain two estimates of the phase space density of stars in the SR. Taking the ratio of these densities gives us an anomaly score $R(x)$ for each star $x$ in the SR. This $R$ value is an estimate of how overdense that star is in the phase space defined by the input features, relative to what would be expected given the distribution of stars in control region (which presumably does not contain a stellar stream, if one exists in the range of proper motions contained in the SR). 

After the ANODE training, we impose fiducial cuts on the data (Section~\ref{sec:datasetfid}) and further divide up the SRs into ROIs (Section~\ref{sec:dataoverview}). We select the 100 highest-$R$ stars in each ROI, which will be used to search for and construct the stellar streams.

The subsequent steps of \vm{} (line finding, protoclustering, streamclustering) are quite different from Paper I, so we will now describe each subsequent step in more detail. 

\subsection{Line-finding}
\label{sec:linefinding}

The next step is to look within the 100 most anomalous stars of each ROI for line-like features that would be the signature of a stellar stream. As in Paper I, we use the Hough transform \citep{Hough:1959qva,Duda:1972ymn}  
to identify lines within the angular positions of these stars, and to provide a parameter that defines the significance of the line. As part of our improvements to \vm{}, we modify our definition of this significance parameter from Paper I.

\begin{figure*}
\begin{centering}
\includegraphics[width=1.9\columnwidth]{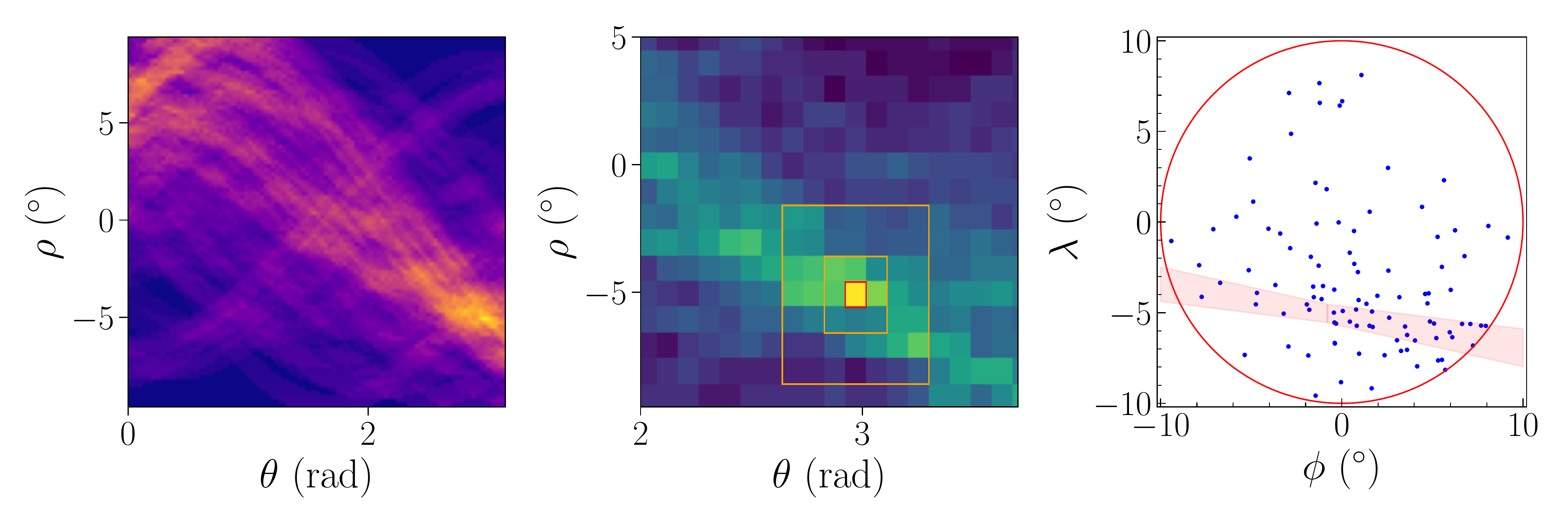}
\caption{Left: all the counts in binned Hough space, for an ROI that belongs to Jhelum. The intensity of each pixel corresponds to the number of stars that lie within
a corresponding box in Hough space, defined by width $\Delta i=5$ and $\Delta j=3$, normalized to the length of the corresponding line segment in the patch. 
Middle: the same 2d histogram as left, but zoomed in around the highest significance pixel, and taking only every pixel strided by $\Delta i=5$ and $\Delta j=3$ in the left 2d histogram. The red box shows the highest significance pixel and the orange boxes indicate the $7\times 7$ strided annulus (with the central $3\times 3$ pixels removed) that is used to estimate the background level. Right: the lines that bound the highest significance bin in Hough space.}
\label{fig:line_example}
\end{centering}
\end{figure*}

The Hough transform parametrizes all possible lines through a point in $(\phi,\lambda)$ space in terms of a {\it Hough curve} in {\it Hough space} $(\rho,\theta)$: 
\beq
 \rho = \phi \sin \theta - \lambda \cos \theta. 
\eeq
Here, $\rho$ is the distance of closest approach between the center of the patch $(\phi, \lambda) = (0,0)$ and the line inclined at angle $\theta$ to the $\hat{\phi}$ axis that passes through the point $(\phi,\lambda)$. If many stars lie on the same line, then many Hough curves will pass through the same $(\rho,\theta)$ point. Thus, the Hough transform converts the problem of line-finding into the simpler task of identifying high-density points in the Hough space of $\rho$ and $\theta$.

For each ROI, we bin the Hough space in a $100\times 100$ grid from $|\rho| \le 10^\circ$ and $0\le \theta\le\pi$. We index the bins with $i,j=0,\dots,99$, such that the map between bin number $(i,j)$ and the left edge of the $(\rho,\theta)$ bin is given by $\rho=-10+i/5$ and $\theta=\pi\times j/100$.

Let $N_{ij}$ denote the number of stars whose Hough curves pass through a box in Hough space with corners $(i,j)$ and $(i+\Delta i,j+\Delta j)$. 
Different lines cover varying amounts of the patch, due to their trajectory across the circle. In order to fairly compare the number of stars in each possible line, we normalize $N_{ij}$ by the length of the line across the patch.  

The hyperparameters $\Delta i$ and $\Delta j$ define the thickness of the streams we are sensitive to. In this work, we have picked relatively small values $\Delta i=5$ and $\Delta j=3$ (corresponding to $\Delta\rho=1^\circ$ and $\Delta\theta=0.09$~rad), so we are explicitly searching for narrow streams (akin to GD-1) as well as somewhat broader streams. Streams covering more of an individual patch would require a larger choice for $\Delta i$ and/or $\Delta j$.

To find the most-significant line parameters in the binned Hough space, we use the same overdensity algorithm as in the GC-identification in Section~\ref{sec:GCfinding}, and described in detail in Appendix~\ref{app:annulus}. Here, we use an annulus of width and height of 7 pixels, strided by $\Delta i=5$ and $\Delta j=3$ and masking the inner $3\times 3$ pixels. The stride is so as to not double-count stars in overlapping boxes. The significance of the pixel $i,j$ using the overdensity algorithm is $\sigma_{{\rm line},ij}$ (see Appendix~\ref{app:annulus} for details of the significance calculation). For each ROI, we select the bin with the highest-significance $\sigma_{\rm line}$ as the single ``best'' line candidate.

In Figure~\ref{fig:line_example}, we show an example of the Hough transform applied to high-$R$ stars of an ROI containing the known stream Jhelum \citep{2018ApJ...862..114S}. The large overdensity in the binned Hough space (left panel) indicates that many of the stars in this ROI lie along a single line. The middle panel of Figure~\ref{fig:line_example} is a zoomed-in look at the highest-significance Hough-space bin  and its neighbors, indicating also the annulus in Hough-space used to compute the significance. Finally, the right-hand panel shows the best-fit line converted back to angular position-space, with the range of $\rho$ and $\theta$ values spanned by the highest-significance bin shown.

\subsection{New protoclustering algorithm}
\label{sec:newprotoclustering}

The line-finding step described in Sec.~\ref{sec:linefinding} identifies high-$R$ stars within an ROI that lie along a relatively narrow line in angular position, producing a significance parameter $\sigma_{\rm line}$. As ROIs are overlapping in proper motion, real stellar streams are expected to be identified by the line-finder in multiple ``nearby" ROIs within the same patch.
In order to obtain the full stream, we must combine the high-significance lines from different ROIs, requiring that they have consistent orientations in position-space as well as having similar values in proper motion-space. This combination of multiple self-consistent ROIs within \textit{a single patch} is referred to as a ``protocluster.'' 

Given that there are ${\mathcal O}(10^5)$ ROIs, we need an efficient, automated clustering algorithm to group together ROIs into protoclusters. The algorithm from Paper I combined nearby ROIs through {\it single-linkage clustering}; an ROI was joined to a protocluster if the distance (in line parameters and proper motion space) between the ROI and any ROI within the protocluster was less than some tunable threshold. This resulted in protoclusters composed of a chain of ROIs, each of which is sufficiently close to a neighbor to complete the linkage. However, the net result was a very diffuse assembly in proper motion, something we do not expect for real streams. 

In \vm{}, we shift from single-linkage clustering to an approach based on hierarchical, iterative clustering. At each step of the new clustering algorithm, we have groups of ROIs
\beq\label{eq:ROIgroup}
{\cal C}_i=\{ROI_{a_1(i)},ROI_{a_2(i)},ROI_{a_3(i)},\dots\}
\eeq
The basic idea of the new algorithm is that two groups, ${\cal C}_{i_1}$ and ${\cal C}_{i_2}$, are joined together if their aggregate proper motions are ``close enough" (to be defined below) and if the line finder run on their concatenation returns a higher significance than on each one separately. This new algorithm results in high-significance protoclusters that by-eye are more similar to properties of known streams (e.g. more compact in proper motion space).

We now describe our new protoclustering algorithm in more detail. We begin with some definitions. 
\begin{itemize}
\item We call a group of ROIs ${\cal C}=\{ROI_{1},ROI_{2},ROI_{3},\dots\}$ {\it valid} if the ROIs are fully pairwise independent, in the sense that they came from pairwise distinct SRs. ROIs from different SRs have their stars chosen by different runs of ANODE, so given the stochastic nature of the neural network training, we expect their anomaly scores to be quasi-independent. Therefore, a stream-like structure that appears in multiple independent ROIs is more likely to be real, compared to a stream-like structure that appears in only a single ROI. Conversely, ROIs that come from the same SR are highly correlated, since their stars were chosen by the same run of ANODE. Thus any structure that appears in multiple ROIs derived from the same SR is not necessarily more likely to be real.

\item We define the {\it line significance} of a group of ROIs to be the significance of the line-finder \textit{re-run} on the concatenation of the highest-$R$ stars from each ROI:
\beq
\sigma({\cal C})=\sigma_{\rm line}(ROI_1{\oplus}ROI_2\oplus ROI_3\dots).
\eeq
Note that we do {\it not} delete duplicate stars in this concatenation, we deliberately double count them, as there is information in the result that a particular star has high-$R$ values from independent runs of ANODE. 

\item We define the {\it proper motion distance} between ${\mathcal C}_1$ and ${\mathcal C}_2$ to be
\beq \label{eq:close_enough}
\chi^2_{\mu}({\mathcal C}_1,{\mathcal C}_2)={(\langle\mu_{\lambda}\rangle_{1}-\langle\mu_{\lambda}\rangle_{2})^2\over \sigma_1^2(\mu_{\lambda})+\sigma_2^2(\mu_{\lambda})} +{(\langle\mu_{\phi^*}\rangle_{1}-\langle\mu_{\phi^*}\rangle_{2})^2\over \sigma_1^2(\mu_{\phi^*})+\sigma_2^2(\mu_{\phi^*})}.
\eeq
Here $\langle\,\,\rangle_{1,2}$ denotes the mean  taken over the stars in ${\mathcal C}_{1,2}$ and $\sigma_{1,2}^2$ the variances.

\item Two valid ROI groups ${\mathcal C}_1$ and ${\mathcal C_2}$ are {\it mergeable} if their concatenation is valid, and if first, their line significance grows upon concatenation
\beq
\sigma({\mathcal C}_{i_1}\oplus{\mathcal C}_{i_2})>\sigma({\mathcal C}_{i_1}),\,\, \sigma({\mathcal C}_{i_2})
\eeq
and second they are ``close enough" in proper motion:\footnote{The threshold for $\chi^2_\mu$ was chosen after inspection of known streams, the detection of false streams in the \Galaxia{} data (see Section~\ref{sec:galaxia}), and the by-eye quality of high-significance streams from the \Gaia{} data.  }
\beq\label{eq:closeenough}
\chi^2_{\mu}({\mathcal C}_{i_1},{\mathcal C}_{i_2})<1.
\eeq
 
 \end{itemize}

Now, with all these definitions in hand, we are ready to describe our new protoclustering algorithm. Given a set of valid ROI groups ${\mathcal C}_1$, ${\mathcal C}_2$, \dots in a patch, consider all mergeable pairs.  Among all mergeable pairs, take the one ${\mathcal C}_{i_1}$, ${\mathcal C}_{i_2}$ with the highest significance after merging. Replace this pair in the list of valid ROI groups with a new ROI group $C_{i_1i_2}={\mathcal C}_{i_1}\oplus{\mathcal C}_{i_2}$.
Repeat until there are no more mergeable pairs of valid ROI groups remaining.

\begin{figure}
\begin{centering}
\includegraphics[width=0.95\columnwidth]{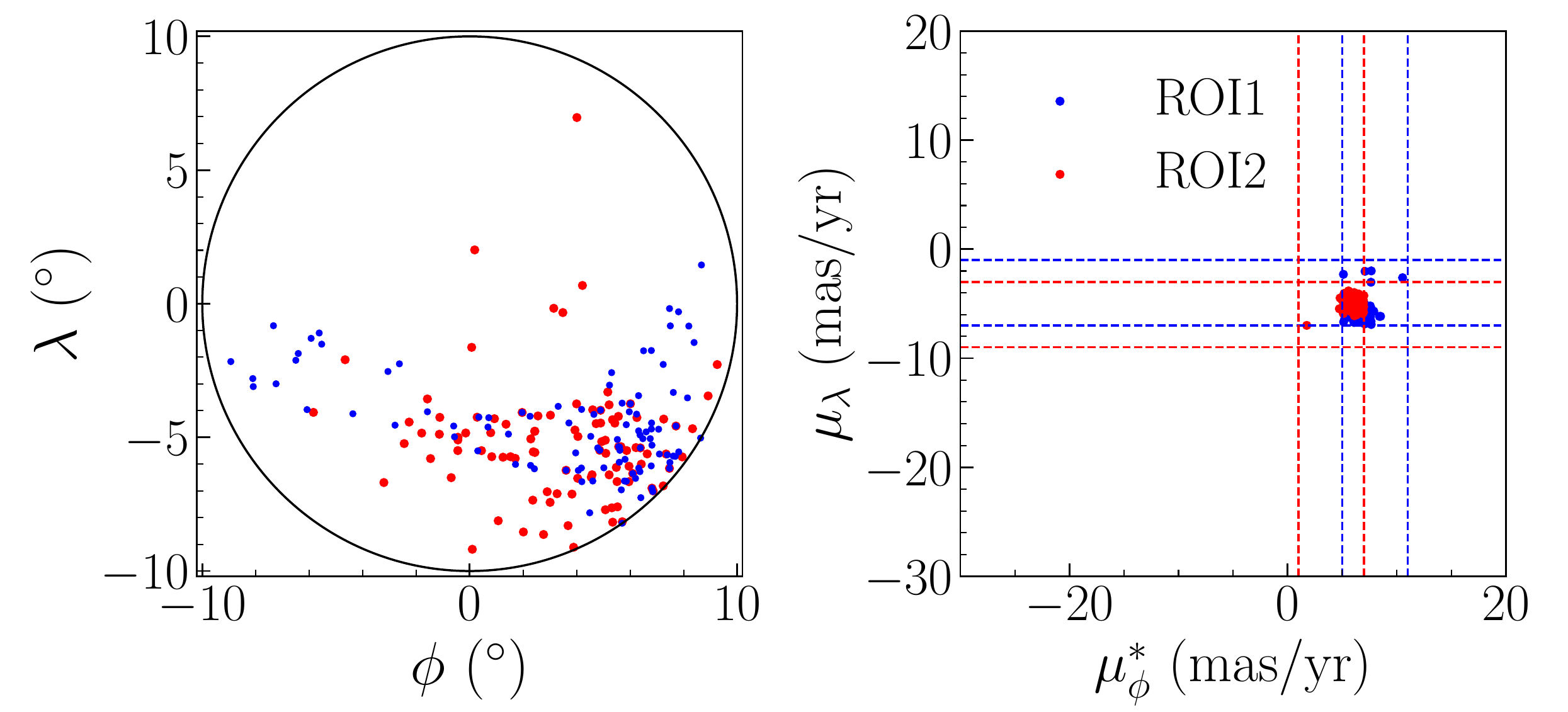}
\caption{Two ROIs from Jhelum that are clustered together into a protocluster, shown in angular position space (left) and proper motion space (right). The proper motion limits defining each ROI are shown in dashed lines with color matching that of the ROI stars (red for ROI1, blue for ROI2)}
\label{fig:pc_example}
\end{centering}
\end{figure}

The algorithm is initialized by defining each individual ROI as its own valid ROI group, i.e.\ ${\mathcal C}_1=\{ROI_1\}$, etc.. After it finishes, one obtains a collection of protoclusters that can be ordered by their significance.  

Figure~\ref{fig:pc_example} shows an example of the above procedure: two ROIs that are clustered together for the real stream Jhelum. $ROI_1$ (blue) is defined by $(\mu_\lambda)_{\rm min}=-7$~mas/yr, $(\mu_{\phi^*})_{\rm min}=5$~mas/yr with the $R$ values derived from ANODE scan over $\mu_{\phi^*}$. $ROI_2$ (red) is defined by $(\mu_\lambda)_{\rm min}=-9$, $(\mu_{\phi^*})_{\rm min}=1$ with the $R$ values derived from ANODE scan over $\mu_\lambda$. Thus these two ROIs are independent as required.  Their line significances are $\sigma(ROI_1) = 6.5$, and $\sigma(ROI_2)  = 6.0$, and their concatenation ${\mathcal C}=\{ROI_1,ROI_2\}$ has increased line significance $\sigma({\mathcal C})  =8.1$. Finally, the proper motion distance between the two ROIs is $\chi_\mu^2(ROI_1,ROI_2)=0.7$.

Through our characterization of the fpr using simulated {\it Gaia}-like observations, to be described in detail in Section~\ref{sec:galaxia}, we are motivated to impose a cut of $\sigma({\cal C})>8$ on the protocluster significance going forward. That is, we will only consider protoclusters with significance greater than this threshold for the subsequent stream-clustering steps of VM2.0, discussed in Sec~\ref{sec:stream_clustering}. Below $\sigma({\cal C})=8$, all of the {\it Gaia} protoclusters are consistent with being false positives, according to both visual inspection and our study of \Galaxia{} protoclusters (see especially Fig.~\ref{fig:pcsig} in Sec~\ref{sec:galaxia}).

\subsection{Merging Duplicate Protoclusters}
\label{sec:duplicatepc}

\begin{figure*}
\begin{centering}
\includegraphics[width=1.9\columnwidth]{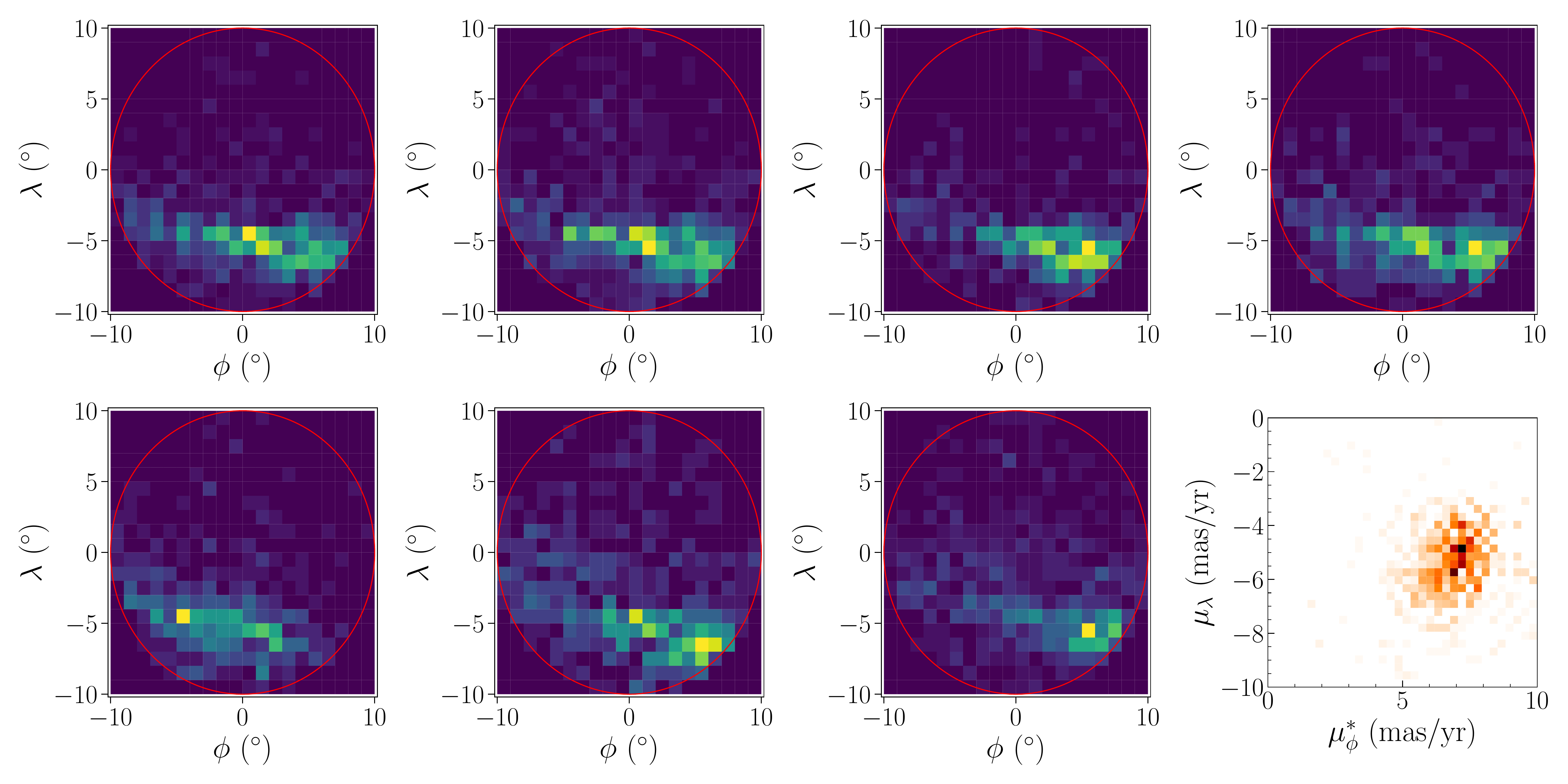}
\caption{An example protostream belonging to Jhelum. This protostream contains 7 duplicate protoclusters. Shown in the last panel is a 2D histogram of the proper motions of all the protocluster line stars (color indicates number density in linear scaling).}
\label{fig:protostream_example}
\end{centering}
\end{figure*}

The requirement that protoclusters be made up of ROIs that are fully pairwise independent means that our protoclustering algorithm can result in multiple duplicate protoclusters corresponding to the same stellar stream within a given patch.
We must therefore develop some set of criteria to decide if two protoclusters ${\cal C}_1$ and ${\cal C}_2$ within the same patch are ``duplicates'' of the same object, and if so, merge them. Since the constituent ROIs of the merged protoclusters are not fully independent of each other, the merging algorithm differs from that used to combine ROIs into independent protoclusters.

Consider two protoclusters ${\cal C}_1$ and ${\cal C}_2$ in a patch.
Each protocluster has a best-fit line associated with it; let the line stars of each be 
${\cal L}_1$ and ${\cal L}_2$ respectively.
${\cal C}_1$ and ${\cal C}_2$ are considered duplicates if ${\cal L}_{1}$ and ${\cal L}_{2}$ overlap above a certain threshold. To quantify this overlap we define
\beq\label{eq:flinedef}
f_{\rm line}=\max\left( {|\{ s\in {\cal L}_2 | s\in {\cal L}_1\}|\over |{\cal L}_2|},\
{|\{ s\in {\cal L}_1 | s\in {\cal L}_2\}|\over |{\cal L}_1|}\right).
\eeq
Here, $|\{s\in {\cal L}_1|s\in {\cal L}_2\}|$ denotes the number of stars in Line 1 which are also in Line 2 (with the obvious extension to $|\{ s\in {\cal L}_2 | s\in {\cal L}_1\}|$), and $|{\cal L}_{1,2}|$ are the number of stars in Lines 1 and 2.
Thus our parameter $f_{\rm line}$ is the fraction of one protocluster's line stars that are found in the other protocluster's set of line stars.\footnote{We note that $|\{s\in {\cal L}_1|s\in {\cal L}_2\}|$ and $|\{s\in {\cal L}_2|s\in {\cal L}_1\}|$ are not necessarily the same, because the line stars can be duplicated through the concatenation of the ROIs.}

Two protoclusters are considered duplicates if 
\beq\label{eq:flinereq}
f_{\rm line}>0.4,
\eeq
i.e., more than 40\% of the line stars of one protocluster are found in the line stars of the other protocluster. The threshold of $40\%$ has been found through testing on high significance protoclusters that are clearly duplicates by eye, especially those corresponding to previously known streams.

Occasionally, two protoclusters are clearly duplicates, by visual inspection of their highest-$R$ stars, but the line finding algorithm results in different best-fit lines that do not satisfy Eq.~\eqref{eq:flinereq}. 
To account for this, apart from the overlap in total number of stars explicitly described in Eq.~\eqref{eq:flinedef}, we also measure the overlap fraction of the highest-$R$ stars, via
\beq\label{eq:fRdef}
f_{ {\rm highest}-R}=\max\left( {|\{ s\in {\cal C}_2 | s\in {\cal C}_1\}|\over |{\cal C}_2|},\
{|\{ s\in {\cal C}_1 | s\in {\cal C}_2\}|\over |{\cal C}_1|}\right), 
\eeq
and we consider two protoclusters as duplicates if
\beq\label{eq:fRreq}
f_{{\rm highest}-R}>0.65,
\eeq
where again, this threshold was set after extensive tests on known streams and other high-significance protoclusters that are clearly duplicates by eye.

Using these two criteria, $f_{\rm line} > 0.4$ OR $f_{{\rm highest-}R}>0.65$, we merge all the protoclusters in a patch into {\it protostreams}. A protostream ${\cal P}$ is the union of a group of duplicate protoclusters: ${\cal P}=\{{\cal C}_1,\, {\cal C}_2,\, \dots\}$ which satisfy the two overlap-fraction criteria using simple single-linkage clustering, i.e.\ a protocluster ${\cal C}_i$ is merged into an existing protostream if it passes the overlap-fraction criteria with {\it any single} protocluster within the protostream.

We now add one last criterion to our merging algorithm: Our overlap-fraction criteria (especially that of Eq.~\eqref{eq:fRreq}) 
can sometimes result in constituent protoclusters in a protostream whose best-fit lines disagree visually; the protoclusters are not well-aligned. If we denote the highest significance protocluster in a protostream by ${\cal C}_1$, we then require all other protoclusters ${\cal C}_i\in{\cal P}$ to satisfy:
\beq\label{eq:protocluster_requirement}
\left| \theta({\cal C}_1) - \theta({\cal C}_i) \right| < 9^\circ, {\rm and}~ \left| \rho({\cal C}_1) - \rho({\cal C}_i) \right| < 1^\circ,
\eeq
where $\theta$ and $\rho$ are the Hough transform parameters found, discussed in Section~\ref{sec:linefinding}.
Any protocluster in ${\cal P}$ that does not satisfy the requirements given in Eq~\eqref{eq:protocluster_requirement} is {\it deleted} from the protostream. 
It is justified to remove these lower significance protoclusters as these are two protoclusters with significant overlap at the level of individual stars, but their best-fit lines are discrepant, so they are not likely to be both correct. 

After this last step, the resulting set of protoclusters in a protostream all have best fit lines that agree visually with one another, and we can take the spread of these lines as an approximate measure of ``uncertainty'' on the protostream itself. We define the significance of a protostream to be the significance of its highest-significance protocluster, 
\beq \label{eq:signficance_protocluster}
\sigma({\cal P})=\max\{\sigma({\cal C}_1),\, \sigma({\cal C}_2),\dots\}.
\eeq 

As an example, we consider a protostream that contains Jhelum (the same ROIs and protocluster considered in the previous subsection). This protostream contains seven duplicate protoclusters with max significance $\sigma({\cal C}) = 15.2$. These are shown in Figure~\ref{fig:protostream_example} as individual 2D histograms, as well as a single 2D histogram of the proper motions of all their line stars. As can be seen, our merging algorithm correctly identifies these seven duplicate protoclusters as corresponding to the same underlying object (Jhelum). The seven duplicate protoclusters have an average proper motion dispersion of $\bar\sigma_{\mu_{\alpha^*}}=1.04$~mas/yr and $\bar\sigma_{\mu_\delta}=1.05$~mas/yr, which is in good agreement with the proper motion dispersions ($0.7-1.2$~mas/yr) reported in~\cite{2019ApJ...881L..37B}.

\subsection{New stream clustering algorithm}
\label{sec:stream_clustering}

The final step of \vm{}, after obtaining the unique protostreams in each patch, is to link the protostreams of {\it different} patches in the sky to produce connected stream candidates. We have made many modifications to the stream clustering algorithm relative to Paper I to improve the quality of the stream candidates (according to a by-eye test) and to reduce the fpr as measured by a scan using synthetic \Gaia{} observations of a smooth simulated Milky Way (described in detail in Section~\ref{sec:galaxia}). We now describe the newly adapted stream clustering algorithm.

\begin{figure}
\begin{centering}
\includegraphics[width=.9\columnwidth]{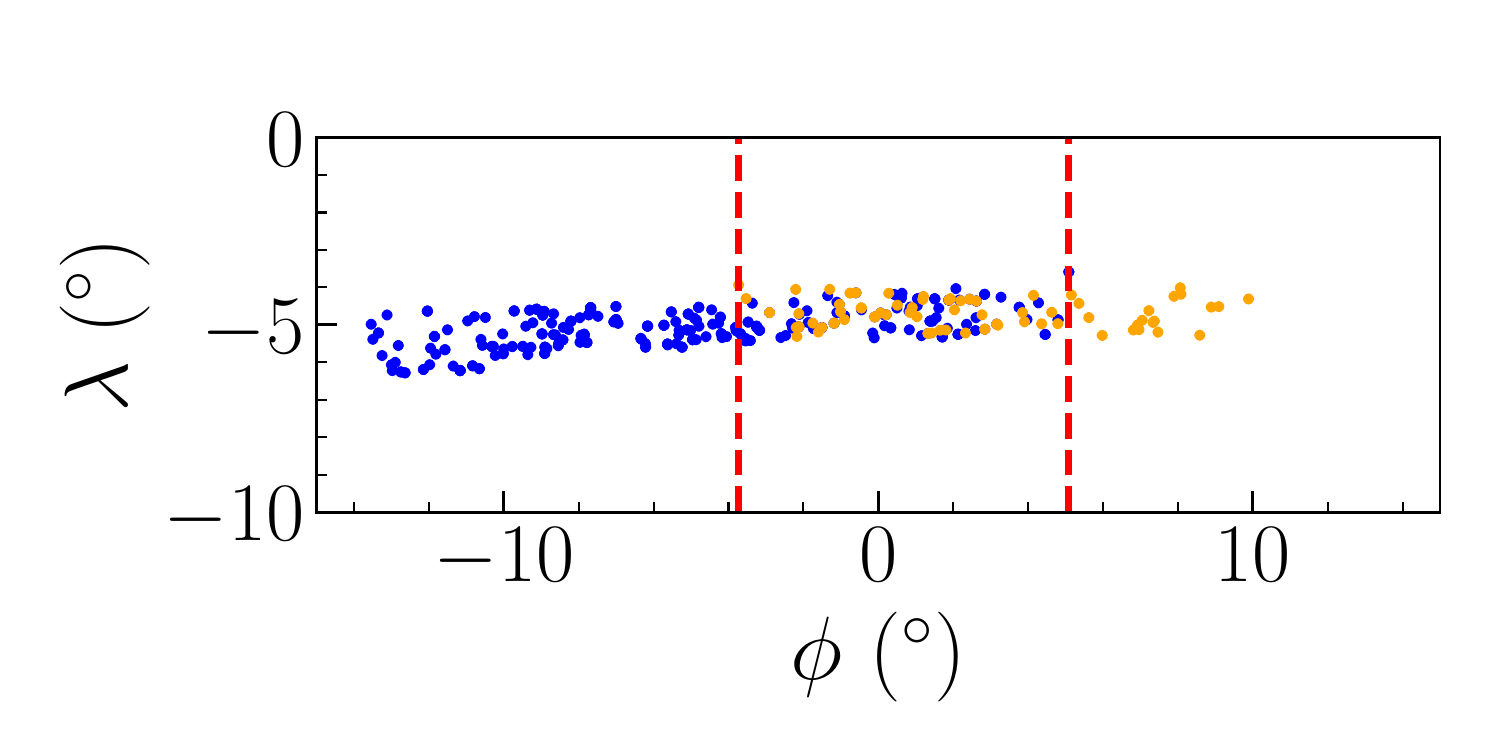}
\caption{An example illustrating the protostream merging criteria, again for Jhelum. Shown in blue and orange are protocluster line stars from two protostreams belonging to two separate patches. The stars are shown in local coordinates centered on the midpoint between the patch centers and rotated so the line stars are aligned along the $x$-direction. 
The overlap region between the protocluster stars is indicated by the red dashed vertical lines.  
}
\label{fig:stream_example}
\end{centering}
\end{figure}

To merge protostreams, we must determine whether they agree in both proper motion and line direction, similarly to the metrics previously developed for the merging of protoclusters. Given a pair of protostreams ${\mathcal P}^{(1)}=\{{\mathcal C}^{(1)}_1,{\mathcal C}^{(1)}_{2},\dots\}$ and ${\mathcal P}^{(2)}=\{{\mathcal C}^{(2)}_1,{\mathcal C}^{(2)}_{2},\dots\}$, we consider every pair of constituent protoclusters ${\mathcal C}^{(1)}_i$, ${\mathcal C}^{(2)}_j$.
We first transform them to local patch coordinates centered on their average sky position. We also rotate the coordinate frame so that their line stars are aligned along the $x$ direction. Finally, we define the overlap region of ${\mathcal C}^{(1)}_i$ and ${\mathcal C}^{(2)}_j$ to be the common extent on the $x$-axis where both protoclusters have stars. See Figure~\ref{fig:stream_example} for an explicit example of this setup, for two high-significance protostreams from Jhelum that are merged together by our criteria.

\begin{figure*}
\begin{centering}
\includegraphics[width=1.85\columnwidth]{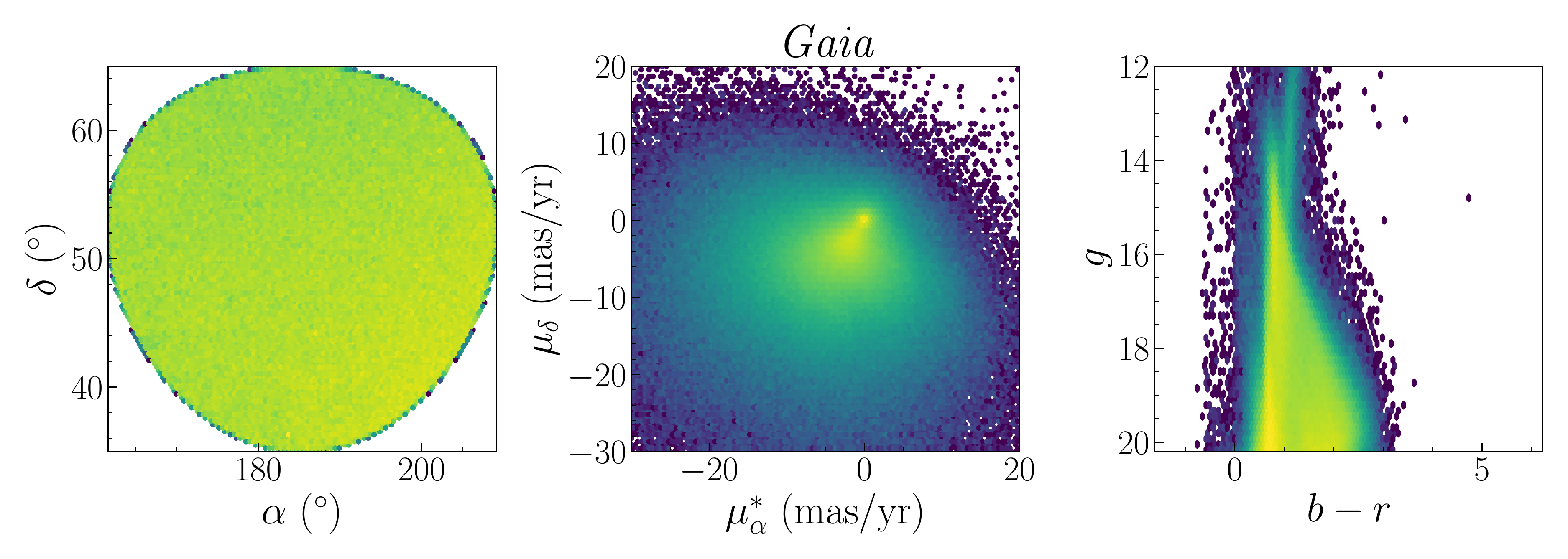}
\includegraphics[width=1.85\columnwidth]{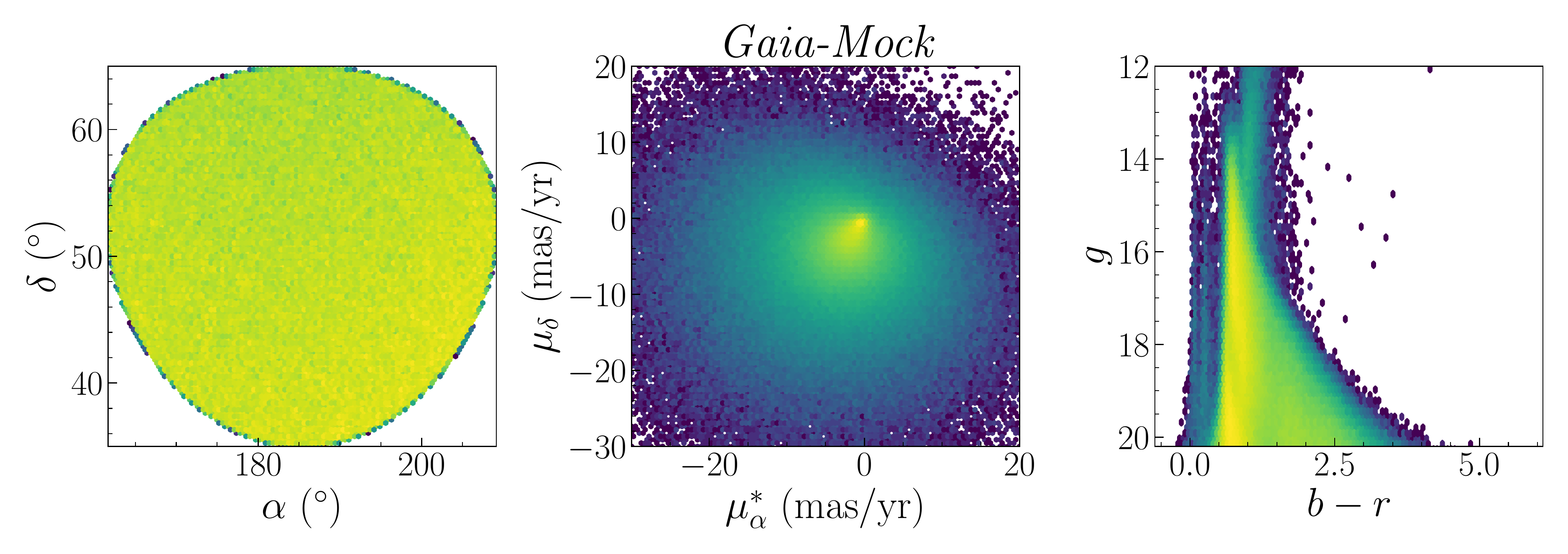}
\caption{Comparison of the position (left column), proper motion (center column) and color-magnitude (right column) of \Gaia{} DR2 (top row) and \Galaxia{} (bottom row) patches centered on $(\alpha,\delta) = (185.4^\circ,50.0^\circ)$. No fiducial cuts have been applied to the data.}
\label{fig:gaia_galaxia_comparison}
\end{centering}
\end{figure*}

If the overlap region is at least $3^\circ$ and the line stars of ${\mathcal C}^{(1)}_i$ and ${\mathcal C}^{(2)}_j$ in the overlap region satisfy $f_{\rm line}>0.4$, then we compute the difference in $\theta$ and $\rho$ between the line stars, call this $\Delta\theta_{ij}$ and $\Delta\rho_{ij}$. The two protostreams are merged into the same stream candidate, if $\Delta\theta_{ij}$ and $\Delta\rho_{ij}$, averaged over all such pairs of protoclusters, satisfy the requirements $|\langle\Delta\theta_{ij}\rangle|<9^\circ$ and $|\langle\Delta\rho_{ij}\rangle|<1^\circ$. (This is the same requirement for visual agreement that we used to clean up protostreams, see Eq.~(\ref{eq:protocluster_requirement}).)

The stream clustering then proceeds via a simple single-linkage clustering, i.e.\ stream candidates ${\cal S}_1=\{{\cal P}_{11},{\cal P}_{12},\dots\}$ and ${\cal S}_2=\{{\cal P}_{21},{\cal P}_{22},\dots\}$ are clustered together into a single stream candidate if any pair of protostreams in them satisfy the above requirements.

Finally, we define the significance of a stream to be the sum-in-quadrature of its constituent protostream significances, which were given by Eq.~\eqref{eq:signficance_protocluster}
\beq
\sigma({\cal S})^2=\sum_{{\cal P}\in {\cal S}} \sigma({\cal P})^2.
\eeq

The final result of this stream clustering algorithm is to smoothly connect stream fragments in a way that allows for the direction of the stream on the sky and for proper motion values to gradually change as one moves along the stream.

\section{Galaxia analysis: estimating the false positive rate}
\label{sec:galaxia}

As the preceding sections indicate, the \vm{} algorithm is a 
 multi-step process, with a large number of ``hyperparameters'' selected to target specific classes of cold stellar streams that are spatially narrow. 
 (For example, the narrowness condition is enforced by our choices for the size of the inner region and outer annulus used to identify the most-significant line in the Hough space, see Section~\ref{sec:linefinding} for details.) 
Given this complexity, it is important to investigate the behavior of \vm{}, in particular with regards to the fpr, i.e., how often does \vm{} identify a stellar stream candidate when none exists?

To provide a benchmark for studies of false positives, we turn to the \Gaia{} DR2 mock catalog described in  \citep{2018PASP..130g4101R}.\footnote{The simulated data is available through the German Astrophysical Virtual Observatory website \url{https://dc.zah.uni-heidelberg.de/browse/gdr2mock/q}.}  
This set of mock observations was generated using the \textsc{Galaxia} code \citep{2011ApJ...730....3S},
and we will refer to it as \Galaxia{} throughout.
Simulated stars within the \Galaxia{} are drawn from a semi-analytic Besan\c{c}on model \citep{2003A&A...409..523R} of the Milky Way, a combination of smooth distributions of stars representing the Milky Way's thin and thick disks, halo, and bulge.  The stellar population models are tuned to approximate the known properties of Galaxy's disks, bulge, and halo components. 
The mock observations of these stars are passed through a 3D dust extinction model based on the Milky Way \citep{2016ApJ...818..130B}, and smeared by measurement errors derived from the nominal \Gaia{} DR2 error model \citep{2005ESASP.576...67D}.

In Figure~\ref{fig:gaia_galaxia_comparison}, we show the distribution of stars in a randomly-selected example patch (centered on $\alpha=185.4^\circ$, $\delta = 50.0^\circ$) from \Gaia{} DR2 (top) and the same patch in \Galaxia{}. There are 896,912 stars in the \Gaia{} patch and 667,426 in the \Galaxia{} equivalent. While differences in the color-magnitude diagrams are visible by eye, the position and proper motion plots are broadly similar.

The smoothness of the underlying model ensures that \Galaxia{} does not have any substructure on the scale of stellar streams. 
Any streams detected by \vm{} in the \Galaxia{} catalog are therefore false positives. Assuming that the \Galaxia{} is modeling the larger-scale features of the \Gaia{} data sufficiently well, we should expect similar numbers of spurious stream candidates created by dust occlusion, motion of disk stars, and large-scale correlations of stellar motion within a patch in both \Gaia{} and \Galaxia. We therefore propose to use the \Galaxia{} to estimate the fpr of stream detections in \Gaia{}.\footnote{Synthetic \Gaia{} catalogs  based on state-of-the-art $N$-body+hydrodynamical simulations also exist, see e.g.\ Ananke \citep{2020ApJS..246....6S}, which is based on FIRE \citep{2015MNRAS.450...53H,Wetzel:2016,Hopkins:2018} and Aurigaia \citep{2018MNRAS.481.1726G} based on Auriga \citep{2017MNRAS.467..179G}. However, the way these mock catalogs  ``upsampled" the initial simulated star particles into synthetic stars led to residual correlations that caused \vm{} to fail already at the ANODE overdensity-finding step. We note that normalizing flows can generate upsampled populations without these unphysical clumping effects \citep{2022arXiv221111765L}, which may allow future stream-finding studies to test methodology using fully cosmological simulations of galaxies.}

To construct the false positive streams in \Galaxia{}, we ran the entire \vm{} algorithm on a quarter of the \Galaxia{} sky, as indicated by the red dots in Figure~\ref{fig:regions}. (Computational limitations prevented us from running \vm{} on a larger portion of the \Galaxia{}.)
Other than sky coverage, the \vm{} analysis steps for the \Galaxia{} data are identical in all respects to those for \Gaia. By comparing the number of stream candidates above a given significance in  \Galaxia{} and \Gaia{} (properly rescaling for the difference in sky coverage), we obtain an estimate of the fpr in \Gaia{}.

\subsection{Initial comparisons -- no additional cuts}

\begin{figure}
\begin{centering}
\includegraphics[width=0.9\columnwidth]{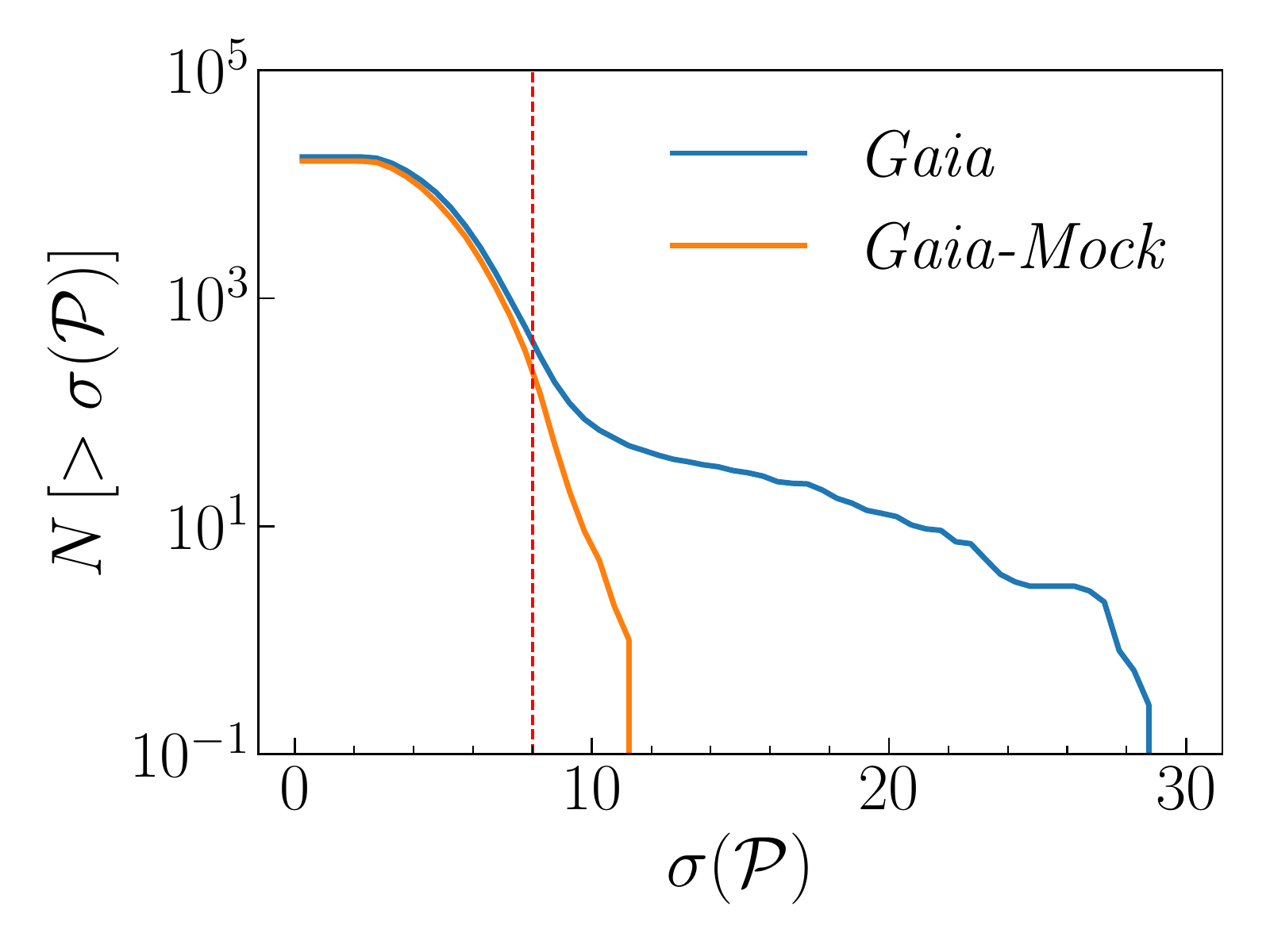}
\caption{
The cumulative number of protoclusters with significance greater than the threshold indicated on the $x$-axis, in  \Gaia{} and \Galaxia{}. We see that they start to diverge strongly after $\sigma({\cal P})\sim 8$, which is indicated by the vertical red dashed line. 
}
\label{fig:pcsig}
\end{centering}
\end{figure}

\begin{figure*}
\begin{centering}
\includegraphics[width=0.5\columnwidth]{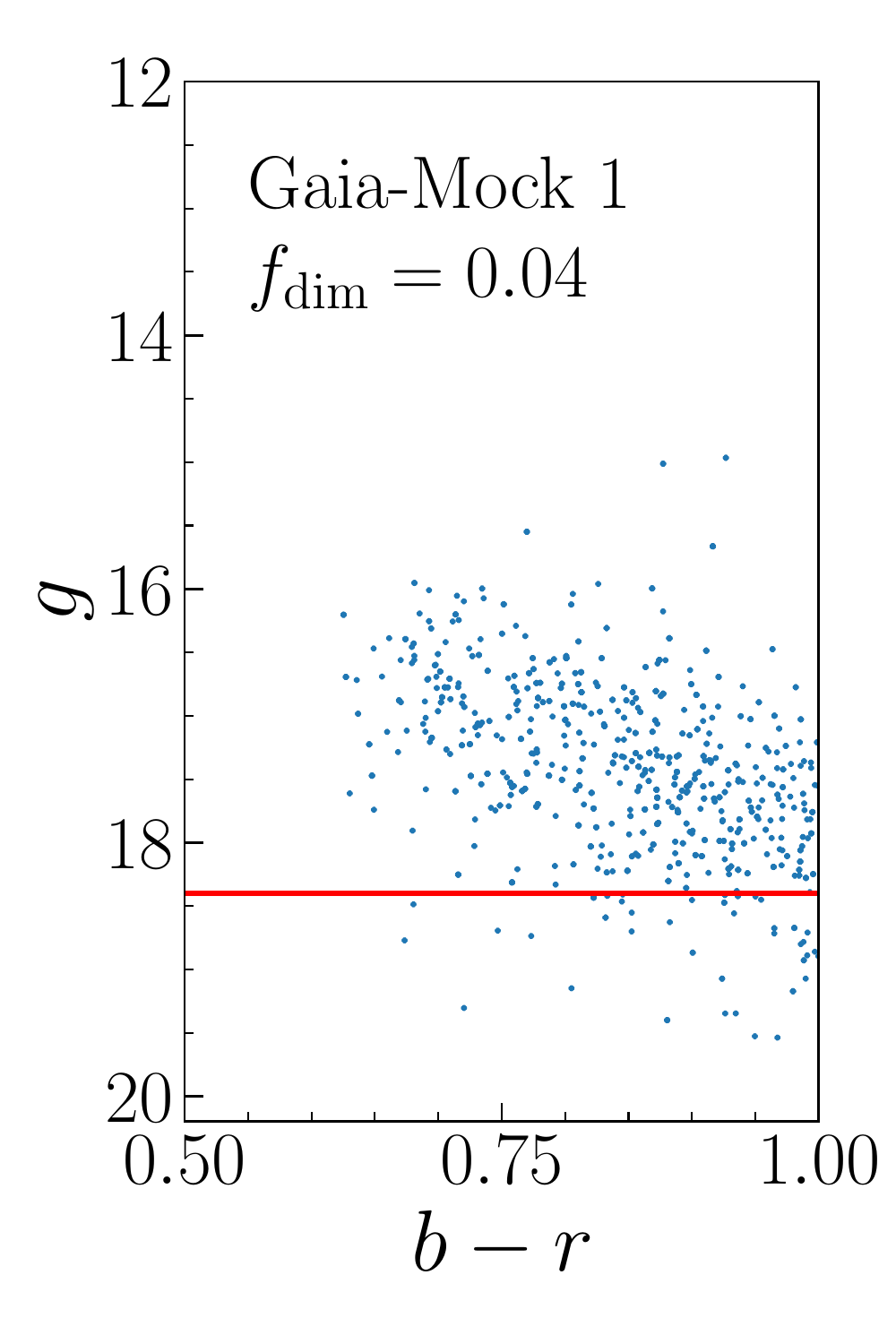}
\includegraphics[width=0.5\columnwidth]{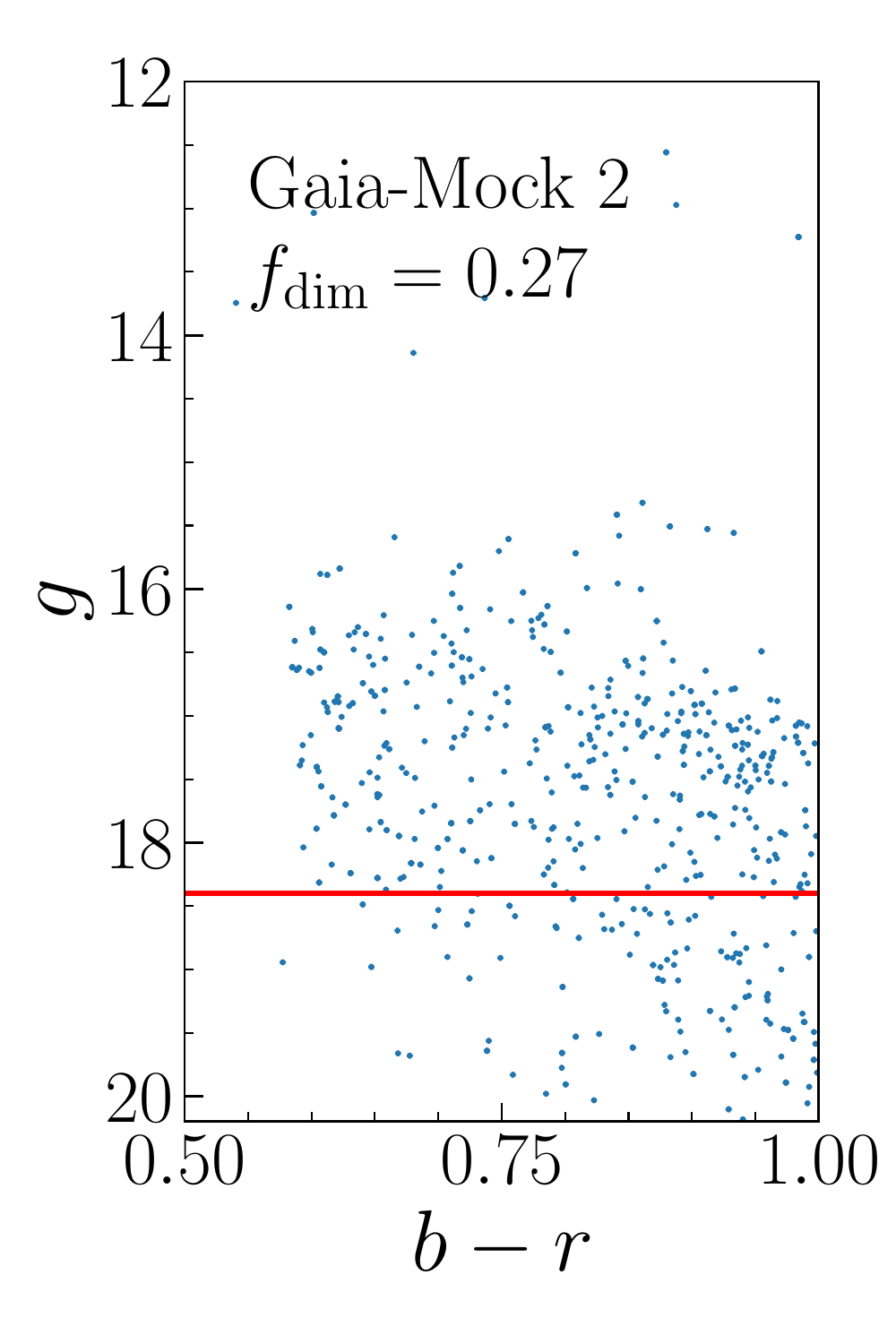}
\includegraphics[width=0.5\columnwidth]{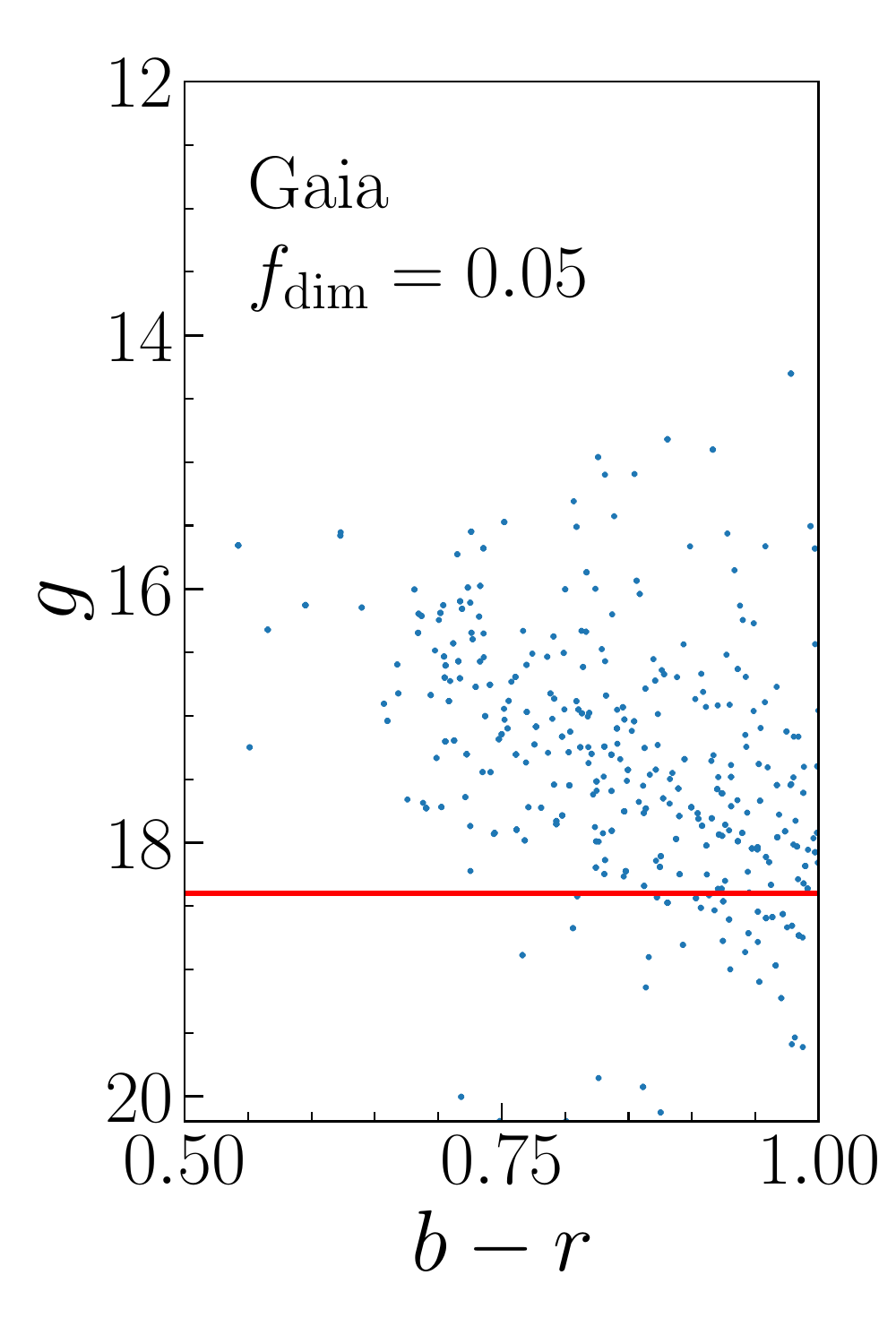}
\includegraphics[width=0.5\columnwidth]{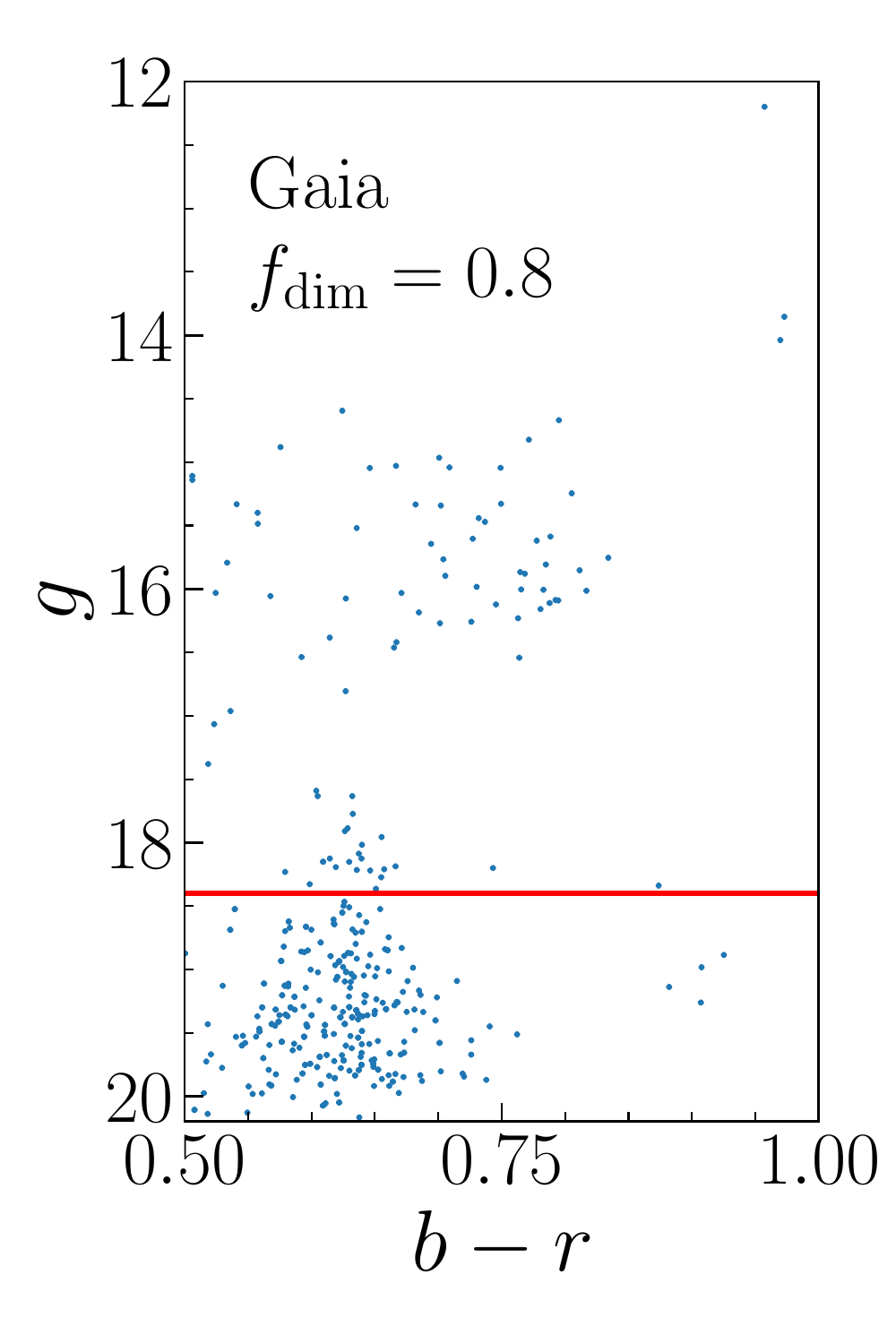}
\caption{From left to right: the color and magnitude of the stars in the two highest-significance {\it Galaxia} protostreams, in a representative \Gaia{} protostream with low $f_{\rm dim}$, and in a representative \Gaia{} protostream with high $f_{\rm dim}$. We see that the first three all have a predominance of bright stars, delineated by the red line at magnitude 18.4.}
\label{fig:galaxiacolormag}
\end{centering}
\end{figure*}

Initially, there are 65,220 \Galaxia{} ROIs within the 44 patches, which should be compared with the 256,557 \Gaia{} ROIs across all 163 patches. After rescaling the former by $163/44$, these values agree to within $\approx 5\%$, showing that \Galaxia{} is an excellent match for the kinematic properties of the real sky.\footnote{The numbers are not exactly the same, because (as in Paper I), ROIs are required to have at least 200 stars in order to be considered in our analysis. So this cut has slightly different efficiency in \Gaia{}  and \Galaxia{}.} In the following, we will quote the counts of ROIs, protoclusters, protostreams and streams from \Galaxia{}, rescaled by 163/44 to match  the number of \Gaia{} patches.

Given that some of the \Gaia{} ROIs are deleted due to GC's, we choose to delete the exact same ROIs from \Galaxia{} for an even comparison (even though \Galaxia{} does not contain any GCs). This is crucial for an accurate estimate of the fpr. After the GC removal, there are 195,344 valid \Galaxia{} ROIs rescaled to 163 patches, compared to 213,874 \Gaia{} ROIs. The numbers of ROIs across both data sets are within 10\% of one another, indicating continued good agreement between \Gaia{} and \Galaxia{}.

After the protoclustering step, there are 59,373 rescaled number of protoclusters in the \Galaxia{} patches, compared to 64,495 protoclusters in all the \Gaia{} patches. These are again within 10\% of one another.  

Shown in Figure~\ref{fig:pcsig} are the number of protoclusters vs.\
protocluster significances for \Gaia{} and \Galaxia{}. We see that the bulk of the \Gaia{} 
distributions are reproduced well by the  \Galaxia{} scan. However,  above $\sigma_{\rm protocluster}\sim 8$, the \Gaia{} and \Galaxia{} distributions appear to diverge in Figure~\ref{fig:pcsig}. 
As discussed above in Section \ref{sec:newprotoclustering}, protoclusters with $\sigma_{\rm protocluster}< 8$ are not visible by eye, and so -- following our philosophy that each protocluster should be a strong detection of a stream fragment -- we cut out such protoclusters. It is reassuring that our comparison with \Galaxia\ confirms that these protoclusters are all likely to be false positives, while above this threshold the proportion of false positives starts to decrease rapidly.

After keeping only protoclusters with $\sigma_{\rm protocluster}> 8$ and merging duplicate protoclusters, we find 848 rescaled number of protostreams in \Galaxia{}. This should be compared to the 1,063 protostreams found in \Gaia{}. While this indicates there could be around 200 protostreams in \Gaia{} corresponding to real streams, our goal was to bring down the fpr. In the next subsection, we will propose one additional cut that will greatly reduce this fpr and leave us with a much more robust sample of candidate streams in \Gaia{}.

\subsection{Cutting on fraction of dim stars}

Given the fpr for protostreams described in the previous subsection, we are motivated to develop additional criteria that we can impose on the protostreams that would improve the fpr. It is important that any cut we devise does not rely on mismodeling differences between \Galaxia{} and the real Milky Way galaxy; i.e., the variables used should be in good agreement between \Galaxia{} and \Gaia. 

From inspection, we observe that many of the \Galaxia{} protostreams are built from stars towards the brighter end of the magnitude range. This is illustrated in Figure~\ref{fig:galaxiacolormag}, which shows the color/magnitude of the stars in the two highest-significance \Galaxia{} protostreams. Such protostreams also occur amongst the \Gaia{} sample; shown in the third panel of Figure~\ref{fig:galaxiacolormag} is a representative \Gaia{} protostream with a similar predominance of bright stars. 

\begin{figure}
\begin{centering}
\includegraphics[width=.9\columnwidth]{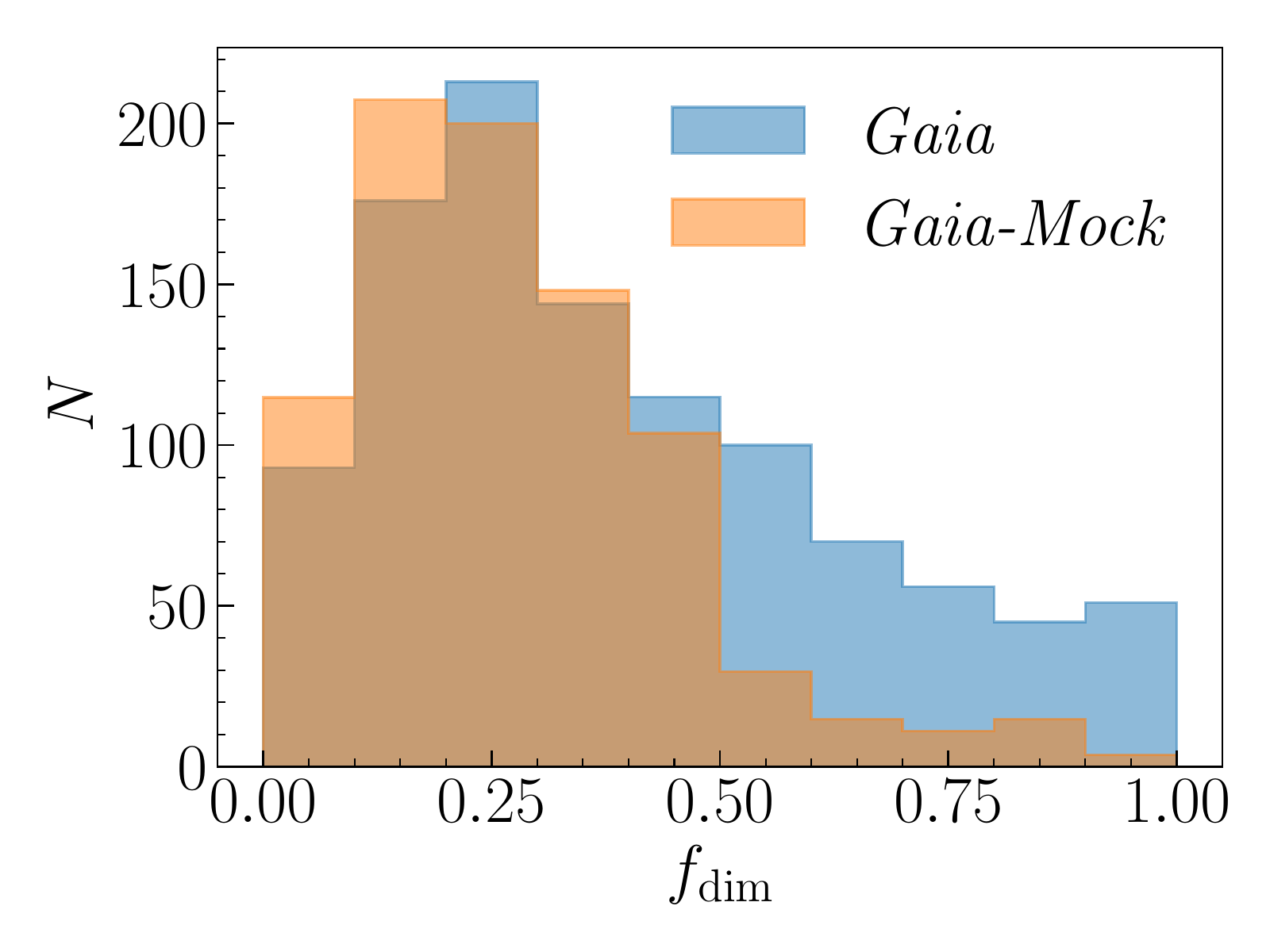}
\caption{The fraction $f_{\rm dim}$ of stars in each protostream with magnitude $g>18.4$. Both \Gaia{} and \Galaxia{} have a similar peak around $f_{\rm dim}\sim 0.2$, but above $f_{\rm dim}\sim0.5$ the number of protostreams (all of which are false positives) in \Galaxia{} is significantly reduced.}
\label{fig:dimfrac}
\end{centering}
\end{figure}

To better identify such protostreams that are mostly composed of brighter stars, we introduce the quantity $f_{\rm dim}$, defined to be the fraction of stars in each protostream with magnitude $g>18.4$. In Figure~\ref{fig:dimfrac}, we show a histogram of the $f_{dim}$ distribution for \Gaia{} and \Galaxia{} protostreams (the latter reweighted to 163 patches, as above). Both \Gaia{} and \Galaxia{} have a similar peak around $f_{\rm dim}\sim 0.2$, indicating that the protostreams dominated by bright stars are a feature of \Gaia{} which is fully reproduced by the \Galaxia{}.  
This strongly suggests that protostreams with large fractions of bright stars (low $f_{dim}$) are more likely to be false positives.\footnote{Whether this is due to particular choices of hyperparameters of our stream-finding algorithm or from ANODE itself is unclear. It is possible that these are all thick-disk stars with poorly-measured parallaxes, such that they were not removed by the fiducial cut described in Section~\ref{sec:datasetfid}.} Meanwhile above $f_{\rm dim}\sim0.5$, the number of protostreams in \Galaxia{} is significantly reduced, but a sizeable number of \Gaia{} protostreams remain.

To reduce the rate of false positives, we will include a quality cut of
\beq\label{eq:fdimcut}
f_{\rm dim}>0.5
\eeq
on the protostreams. After imposing this cut, there are 317 \Gaia{} protostreams but only 70 rescaled number of \Galaxia{} protostreams. In the rightmost panel of Figure~\ref{fig:galaxiacolormag} is shown a representative \Gaia{} protostream that survives Eq.~(\ref{eq:fdimcut}). With these protostreams in hand, we are ready to compare the final step of our algorithm -- stream clustering -- within \Gaia{} and \Galaxia.

\begin{figure}
\begin{centering}
\includegraphics[width=.9\columnwidth]{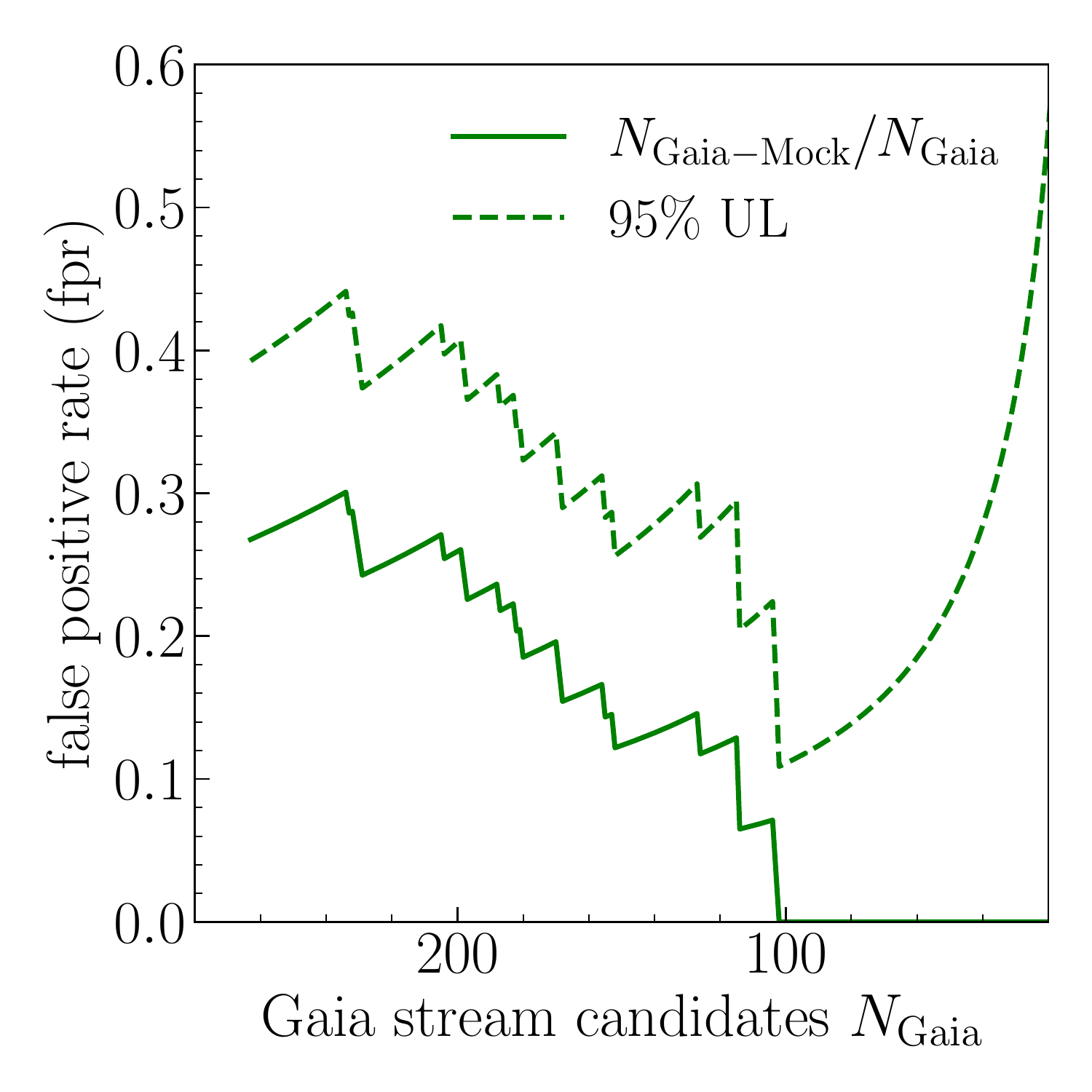}
\caption{The fpr plot for complete stream candidates (i.e.\ after the final stream clustering step). Solid and dashed denote the nominal fpr and the 95\% UL on the fpr assuming Poisson statistics, respectively. }
\label{fig:stream_fpr_cut}
\end{centering}
\end{figure}

\subsection{Streams in \Galaxia}

After clustering the \Galaxia{} protostreams with the $f_{\rm dim}>0.5$ cut in place, we find 70 rescaled number of \Galaxia{} streams, with max significance $\sigma_{\rm stream} = 8.86$. Similar clustering on the \Gaia{} data results in 262 streams over the full sky. Placing the same cut on significance for \Gaia{} and \Galaxia{} streams, we obtain an estimate of the fpr for the streams above that significance.

 \begin{figure*}
\begin{centering}
\includegraphics[width=2\columnwidth]{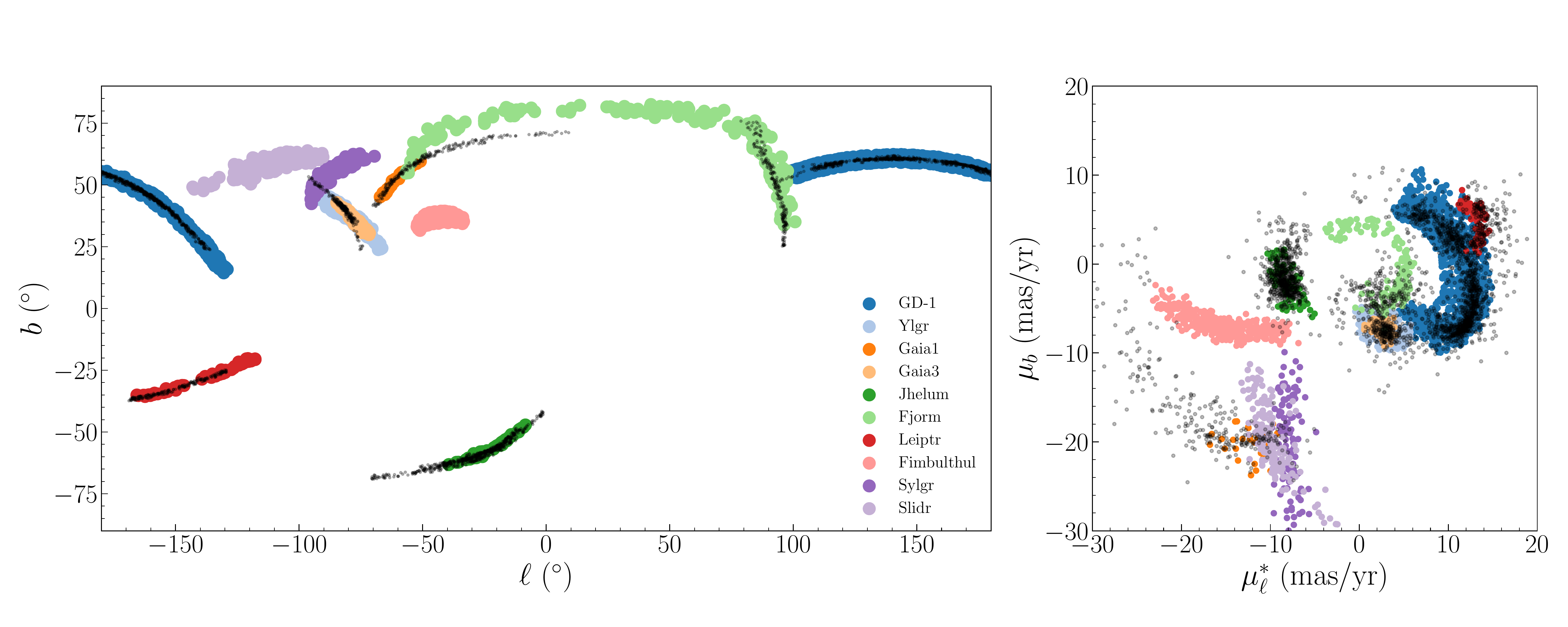}
\caption{Streams discovered or re-discovered in \Gaia{} DR2 data by \streamfinder{}, with position and proper motions digitized from \streamfinder{} papers \protect\citep{2018MNRAS.477.4063M,2018MNRAS.478.3862M}. (GD-1 coordinates from \protect\cite{2018ApJ...863L..20P}.) Overlaid in black are the \via{} stream candidates corresponding to a subset of these known streams. Fimbulthul, Sylgr, and Slidr are not identified in \via{}, see text for details. Left: positions in Galactic coordinates. Right: proper motions in Galactic coordinates.}
\label{fig:streamfinder}
\end{centering}
\end{figure*}

The fpr vs.\ number of \Gaia{} streams above a given significance is shown in Figure~\ref{fig:stream_fpr_cut}. Given the reduced statistics of the \Galaxia{}, we take a conservative approach and adopt the 95\% upper limit (UL) on the fpr, derived from Poisson statistics on the actual (integer) number of \Galaxia{} streams. The 95\% UL on the fpr is minimized at 11\% with a requirement that $\sigma_{\rm stream}>8.86$ -- when considering the 95\% UL, it is not beneficial to cut harder on significance after the last \Galaxia{} stream.  Applying this threshold to the real Milky Way data results in 102 \Gaia{} stream candidates in the full-sky scan (163 patches), of which we expect at least $\sim 90$ of these to correspond to real streams, assuming the \Galaxia-derived fpr holds in the \Gaia{} data.

\section{Streams in Gaia DR2} 
\label{sec:rediscovery}

Having built and tested \vm{}, and quantified its fpr, we now proceed to apply it to \Gaia{} DR2. 
In Section~\ref{sec:gd1}, we focus on \vm's re-identification of the well-known stream GD-1. We treat this stream separately as it has been identified in numerous other analyses, as well as being the focus of the first implementation of \textsc{Via Machinae} in Paper~I. Section~\ref{sec:sagittarius} contains a brief discussion of \vm{} stream candidates that may correspond to fragments of the Sagittarius stream. The present work, being optimized for narrow streams, is not expected to recover all of the Sagittarius stream, given its rather large width. Still, Sagittarius is a very distinctive overdensity in the sky and so it is not surprising that \vm{} picks up some fragments of it.  In Section~\ref{sec:streamfinder}, we consider the stellar streams that have been identified by \streamfinder, another automated method for finding stellar streams within \Gaia{} data that has also been applied over the whole sky \citep{2018MNRAS.477.4063M,2018MNRAS.478.3862M}. Both the \streamfinder{} stellar streams that have and have not been re-identified in \vm{} are of interest, as the intersection and complementarity of the two methods can better our understanding of the methods and their limitations.
Finally, the newly found stream candidates that do not have a previously discovered analog in the literature are briefly presented in Section~\ref{sec:newstreams}. Given that this is still primarily a methodology paper, we leave a more detailed study of the new stream candidates and their broader astrophysical implications to future work.

\subsection{GD-1}
\label{sec:gd1}

\begin{figure*}
\begin{centering}
\includegraphics[width=1.9\columnwidth]{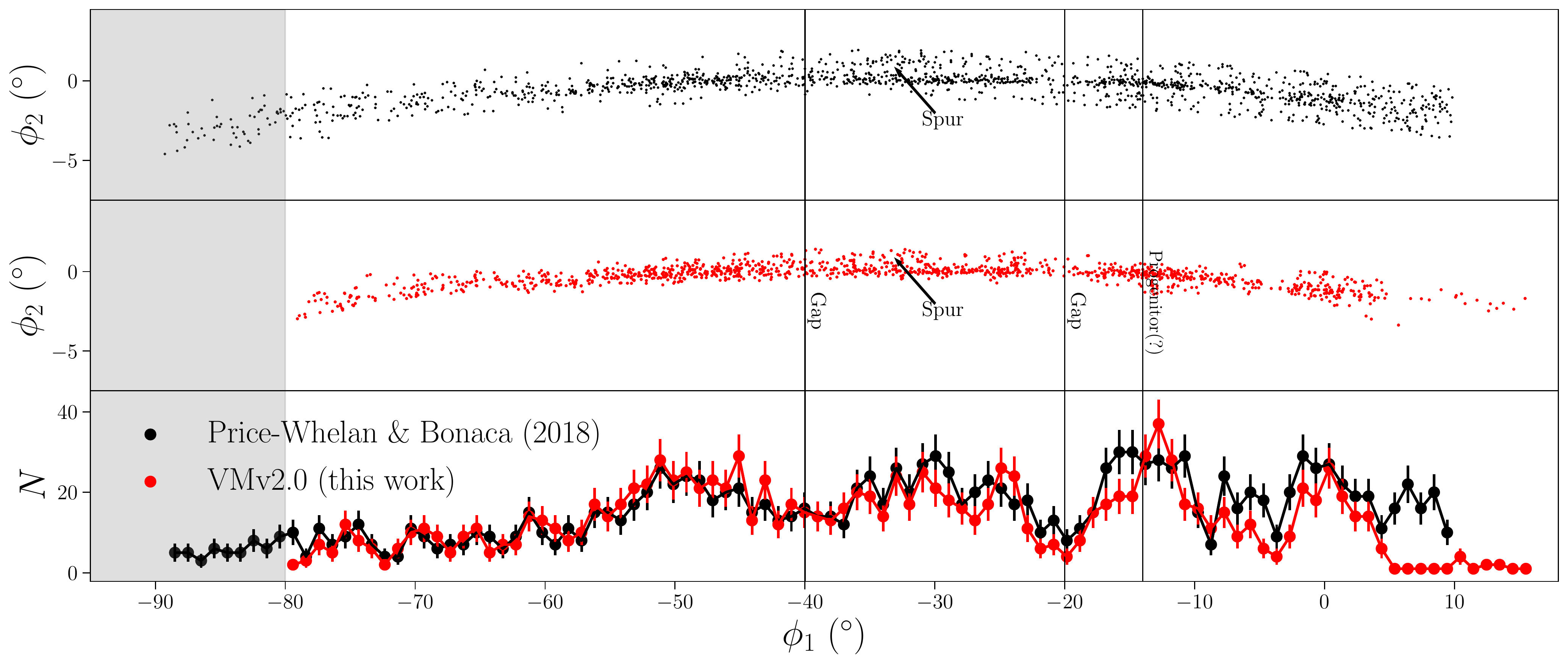}

\caption{Top: positions of stars identified as likely members of GD-1 by \protect\cite{2018ApJ...863L..20P}, using cuts in position, proper motion, and color-magnitude space. Middle: positions of stars identified as likely members of GD-1 by the \vm{} algorithm. Bottom: histogram of both catalogs of GD-1 stars in $1^\circ$-wide bins along the stream coordinate $\phi_1$. Stream coordinate system $\phi_1-\phi_2$ defined in \protect\cite{2010ApJ...712..260K}. The region below $\phi_1\approx-80$ is shaded gray to indicate that it was not included in the 163 patches of the \vm{} scan, due to being too close to the Galactic disk.}
\label{fig:gd1_binned}
\end{centering}
\end{figure*}

\begin{figure*}
\begin{centering}
\includegraphics[width=2\columnwidth]{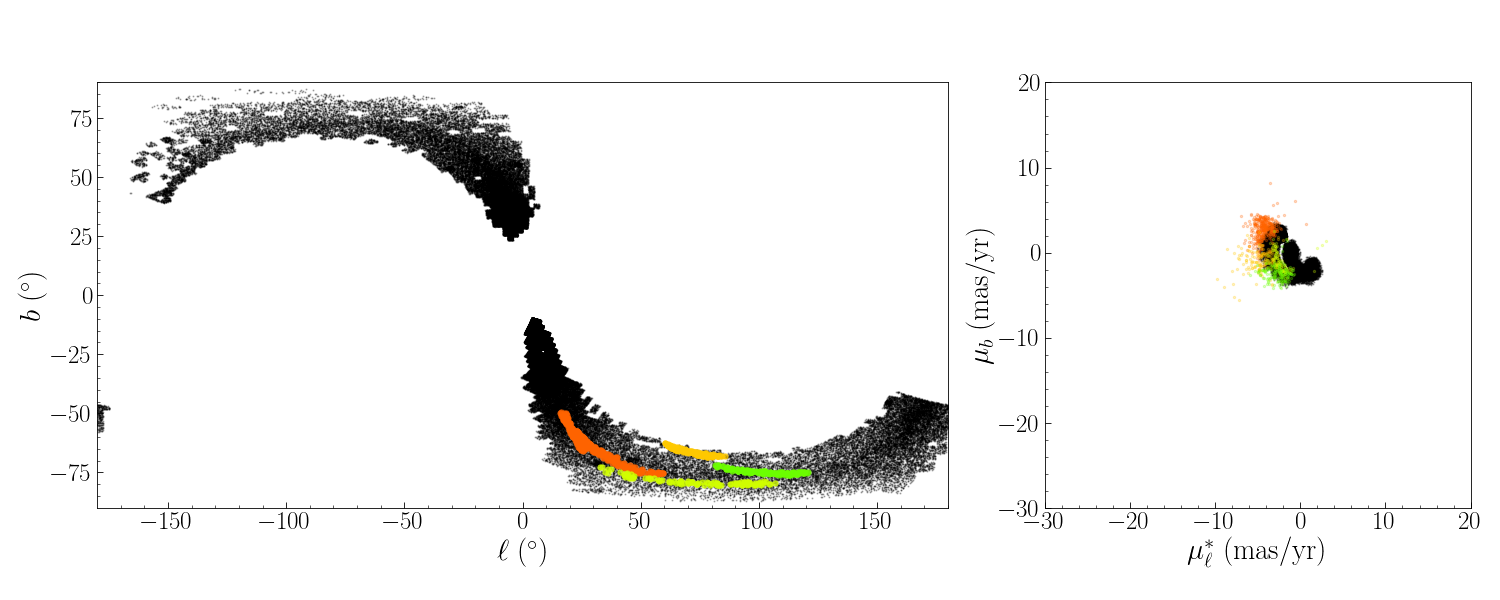}
\caption{The position (left) and proper motion (right) of \Gaia{} DR2 Sagittarius stream stars (black points) overlaid with the four stream candidates from \vm{} which are possible members of Sagittarius. Sagittarius stream stars obtained from \protect\cite{2020AA...635L...3A}.
\label{fig:sag_streams}}
\end{centering}
\end{figure*}

The GD-1 stellar stream served as the test-bed for the first iteration of the \textsc{Via Machinae} algorithm, due to its extent, stellar density, and existing catalogs of likely stellar members. Originally discovered in SDSS data, the stellar membership list has been refined and extended using \Gaia{} data. For a study of the member GD-1 stars produced by \vm, we use as a benchmark \cite{2018ApJ...863L..20P}, which identifies 1,985 stars as likely members of GD-1, based on proper motion, color, magnitude, and position within a defined footprint. 
In \cite{2022MNRAS.509.5992S}, the first iteration of \textsc{Via Machinae} identified 1,688 likely members of GD-1, of which 738 were likewise identified as stream members by \cite{2018ApJ...863L..20P}.

Due to its distinctiveness in all its kinematic, spatial, and photometric features, GD-1 is the highest-significance stream candidate in this updated \vm{} analysis, with $\sigma_{\rm stream} = 83$. We identify 1,252 stars as likely members using \vm{}, of which 820 (65\%) are likewise identified as stream members by \cite{2018ApJ...863L..20P}. As shown in Fig.~\ref{fig:streamfinder}, the reconstructed stream overlays the known path of GD-1 in both angular position and proper motion space.

In Figure~\ref{fig:gd1_binned}, we show the comparison between the \vm{} stars and the existing GD-1 catalog \citep{2018ApJ...863L..20P} in the stream-aligned coordinate system of \cite{2010ApJ...712..260K}.\footnote{The location of the GD-1 stars in proper motion and color/magnitude are shown in Figures~\ref{fig:streamfinder} and Figures~\ref{fig:known_streams_colormag}, respectively.} As can be seen, the \vm{} algorithm reproduces well-known features of GD-1, including the progenitor, the spur, and the two known gaps in the stream. Note that below $\phi_1 \sim -80^\circ$ (indicated by the gray shaded region in Fig.~\ref{fig:gd1_binned}), GD-1 is located near the Galactic disk, and as a result the patches containing this part of the stream were not part of the \vm{} analysis.

Compared to the analysis of GD-1 from Paper I, the new algorithm no longer identifies a secondary stream candidate perpendicular to GD-1. This is a result of our stricter criteria for combining high-significance ROIs into protoclusters and then into protostreams. Our algorithm now identifies more of the stream than in Paper I: in particular, we now capture a section between $\phi_1 \sim [-10^\circ,10^\circ]$, which was missed in our original algorithm due to the single proper motion analysis ($\mu_\lambda$ used in the ANODE step was too close to zero). As described in Section~\ref{sec:dataoverview}, in \vm{}, both proper motion coordinates are used in separate ANODE trainings, allowing for this set of stream stars to be tagged as anomalous using the $\mu_{\phi^*}$ search regions. Despite this improvement, \vm{} continues to miss stars in GD-1 around $\phi_1 \sim 15^\circ$, likely due to both proper motions of these stars being close to zero.

Overall, we re-identify a higher percentage of likely stream members compared to Paper I. The fewer number of stars in our stream compared to Paper~I (along with the tighter clustering around the stream-track) suggests that the new algorithm is successful in combining fragments of the stream cohesively, and rejecting high-significance ROIs that are incompatible with the stream's path across the sky.

\subsection{The Sagittarius Stream}\label{sec:sagittarius}

Based on the stellar membership, at least four of the 102 stream candidates identified \vm{} may be subcomponents of the Sagittarius stream \citep{2002ApJ...569..245N,2003ApJ...599.1082M}.

\begin{table*}
\begin{center}
\begin{tabular}{|c|c|c|c|c|}
\hline 
\textbf{Stream Name} & \textbf{References} & $\mathbf{N}_{\rm SF}$ & $\mathbf{N}_{\vm}$ & \textbf{Status}\\
\hline 
\hline 
Gaia1& M2018b & 31 & 308 & {\bf Found}\\
\hline 
Gaia2& M2018b & - & - &  SRs excluded (GCs) \\
\hline 
Gaia3& M2018b & 57 & 276 &  Same as Ylgr\\
\hline 
Gaia4& M2018b & 8 & - &  Not found (too few stars)\\
\hline 
Gaia5& M2018b & 37 & - &  Not found (too few stars) \\
\hline 
\hline 
Indus&  M2018b, S2019  &  150 & - &  Not found (too dim)\\
\hline 
Jhelum&  M2018b, S2019  & 63  &  665 & {\bf Found}\\
\hline 
Orphan& M2018b, F2019 & - & - &  SRs excluded (pms too small)\\
\hline 
\hline 
Gjoll& I2019 & - & - &  SRs excluded (too few stars)\\
\hline 
Fjorm& I2019 & 148 & 219 &  {\bf Found} \\
\hline 
Leiptr& I2019& 67 & 233 &  {\bf Found}\\
\hline 
Svol&  I2019& 45 & - &  Not found (too few stars) \\
\hline 
Fimbulthul& I2019 & 309 & -  &  Found by ANODE but too wide\\
\hline 
Ylgr& I2019 & 349 & 276 &  {\bf Found}\\
\hline 
Sylgr& I2019 &  103 & - &  Found by ANODE but too wide\\
\hline 
Slidr& I2019 & 156 & - &  Found by ANODE but too wide\\
\hline 
\hline 
Phlegethon& I2018 & - & - &  SRs excluded (too few stars)\\
\hline 
\hline
GD1 & GD2006, PWB2018 & 1985 & 1252 & {\bf Found}\\
\hline 
\end{tabular}
\caption{Streams discovered or re-discovered in \Gaia{} DR2 using the \streamfinder{} algorithm. The references in the second column correspond to the following: M2018b \citep{2018MNRAS.481.3442M}, S2019 \citep{2019ApJ...885....3S}, F2019 \citep{2019MNRAS.486..936F}, I2019 \citep{2019ApJ...872..152I}, I2018 \citep{2018ApJ...865...85I}, GD2006 \citep{2006ApJ...643L..17G}, PWB2018 \citep{2018ApJ...863L..20P}. The columns $N_{\rm{SF}}$ and $N_{\rm{VM2}}$ correspond to the number of stars found by \streamfinder{} and \vm{} respectively.}
\label{tab:knownstreamsSF}
\end{center}
\end{table*}

To identify potential components of Sagittarius, we compare the stars within all our stream candidates with the 294,344 \Gaia{} DR2 stars identified as belonging to the Sagittarius stream in \cite{2020AA...635L...3A}. 
Nine stream candidates have at least one star in the Sagittarius catalog. However five of these appear to not overlap significantly in proper motion space with Sagittarius. We show the four likely subcomponents of Sagittarius in Figure~\ref{fig:sag_streams}. 

Sagittarius itself is far too wide to be reconstructed by the post-ANODE stream-finding steps. However, the member stars of Sagittarius do often have high-$R$ value, and it is therefore not surprising that some subcomponents are identified as stream candidates by the \via{} algorithm. It is perhaps notable that these possible Sagittarius stream segments mostly have significances between $\sim 9-10$ (with one having $\sigma_{\rm stream} = 16$). These are at the lower-end of the stream significances we consider in this work (recall the cutoff from the Galaxia fpr study was $\sigma_{\rm stream}>8.86$). The fact that these less-significant stream candidates still correspond to real objects gives us more confidence that the other 102 streams above the cutoff are also real (and not only the highest-significance ones).

\subsection{Comparison with \streamfinder}
\label{sec:streamfinder}

\begin{figure*}
\begin{centering}
\includegraphics[width=1.6\columnwidth]{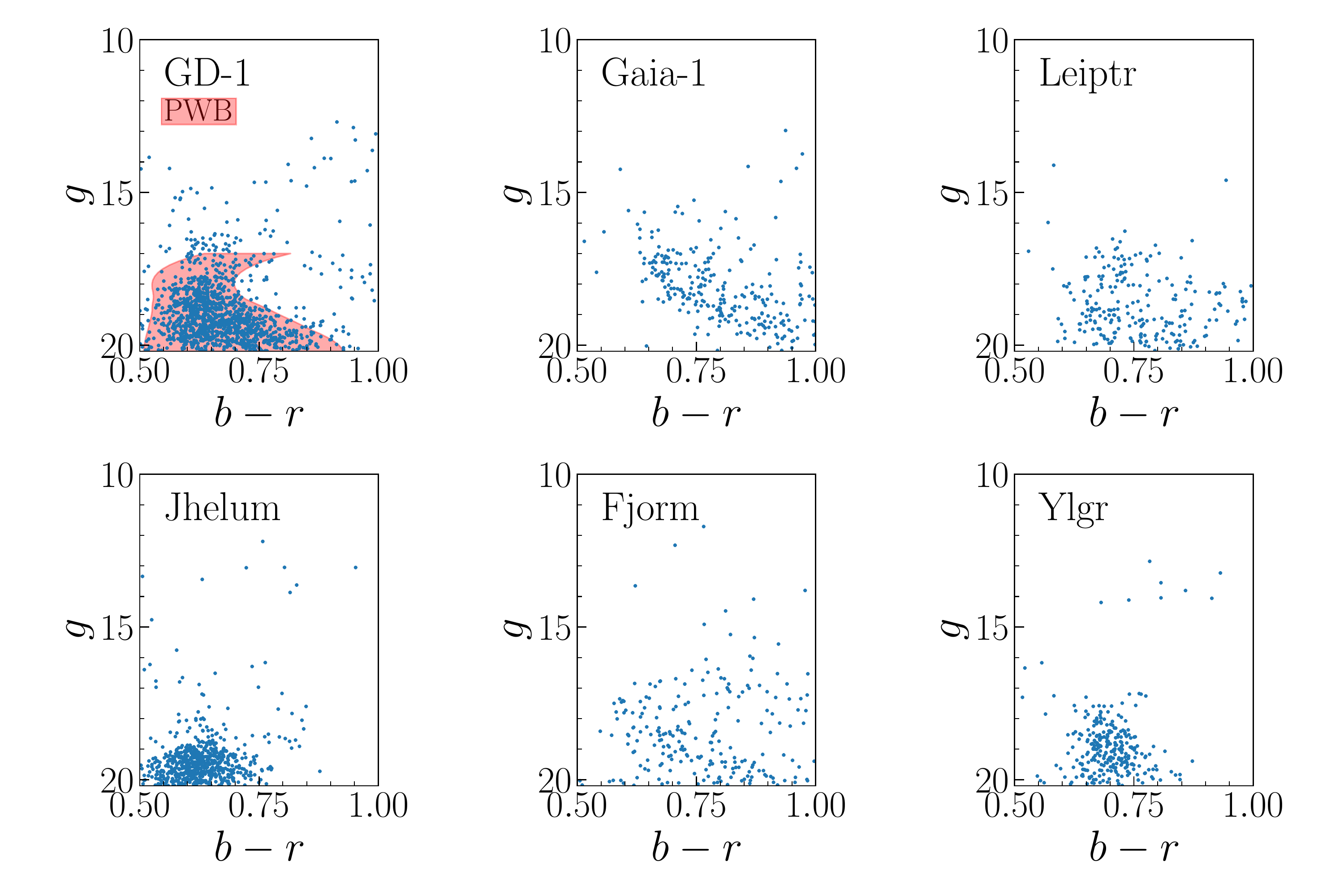}
\caption{Stars in GD-1 and the five streams re-identified by \vm{}, in color $b-r$ versus magnitude $g$ space. The shaded red region shown for GD-1 indicates a (smoothed) envelope enclosing the subset of \Gaia{} stars in our GD-1 catalogue which were identified as likely stream members by \protect\cite{2018ApJ...863L..20P}.
\label{fig:known_streams_colormag}}
\end{centering}
\end{figure*}

Next, we turn to an evaluation of the performance of \vm{} on streams previously identified by the \streamfinder{} algorithm. 
\streamfinder{} is the only other automated search for stellar streams using the \Gaia{} data set, and a comparison enables us to better understand the strengths and limitations of \vm{}. Unlike \vm{}, however, \streamfinder{} assumes a Milky Way potential and searches for stellar streams within defined orbital cones, with photometry following an isochrone. \vm{} remains agnostic to such choices.

In \streamfinder{} survey I, \cite{2018MNRAS.481.3442M} discovered five new stream candidates -- named Gaia-1 through Gaia-5 -- and demonstrated that their method could find four previously-discovered streams (GD-1, originally found in \cite{2006ApJ...643L..17G}, Indus and Jhelum discovered in \cite{2018ApJ...862..114S}, and the Orphan stream found in \cite{2007ApJ...658..337B}). The second survey, \streamfinder{} II  \citep{2019ApJ...872..152I}, discovered eight new stream candidates (Gjoll, Fjorm, Leiptr, Svol, Fimbulthul, Ylgr, Sylgr, Slidr). Finally, \cite{2018ApJ...865...85I} describes a new stream candidate, Phlegethon.\footnote{In this work, we restrict our comparison to the \streamfinder{} surveys that used \Gaia{} DR2, for a fair comparison. \streamfinder{} did subsequently apply their method to (e)DR3 \citep{2020arXiv201205245I}, but we save a comparison to these results for forthcoming work.}

First, we point out that Gaia-3 and Ylgr appear to be the same stream based on their positions and proper motions, which we show in Figure~\ref{fig:streamfinder}. A detailed study of the chemical abundances of members of both streams would be required to confirm this identification. 
Counting these as the same stream yields a total of 17 streams and stream candidates that were discovered or confirmed with \streamfinder\ in \Gaia{} DR2. These stream candidates, their proper motion ranges, and their status in \vm{} are summarized in Table~\ref{tab:knownstreamsSF}. Having already discussed GD-1 in the previous subsection, here we focus on the rest.

Three of the streams, Gjoll, Phlegethon, and Orphan, were excluded from our search because their SRs contained too few stars (Gjoll, Phlegethon) or occurred at proper motions close to zero (Orphan). Additionally, Gaia-2 was excluded from our search because all of its ROIs contained a GC candidate (see the selection requirements for \vm{} in Section~\ref{sec:inputs}).

Based on the cuts that we use in \vm{} (and excluding the previously-discussed GD-1), we are left with 12 previously known stream candidates that fall into regions of the sky that were analyzed by our algorithm.
Of these, we rediscover five using \vm{}: Gaia-1, Gaia-3/Ylgr, Jhelum, Fjorm and Leiptr. Our method did not identify seven candidates: Gaia-4, Gaia-5, Indus, Svol, Fimbulthul, Sylgr, and Slidr. We now discuss both sets of streams in turn, beginning with the streams that are found in \Gaia{} data by both \streamfinder{} and \via{}.

\begin{figure*}
\begin{centering}
\includegraphics[width=1\columnwidth]{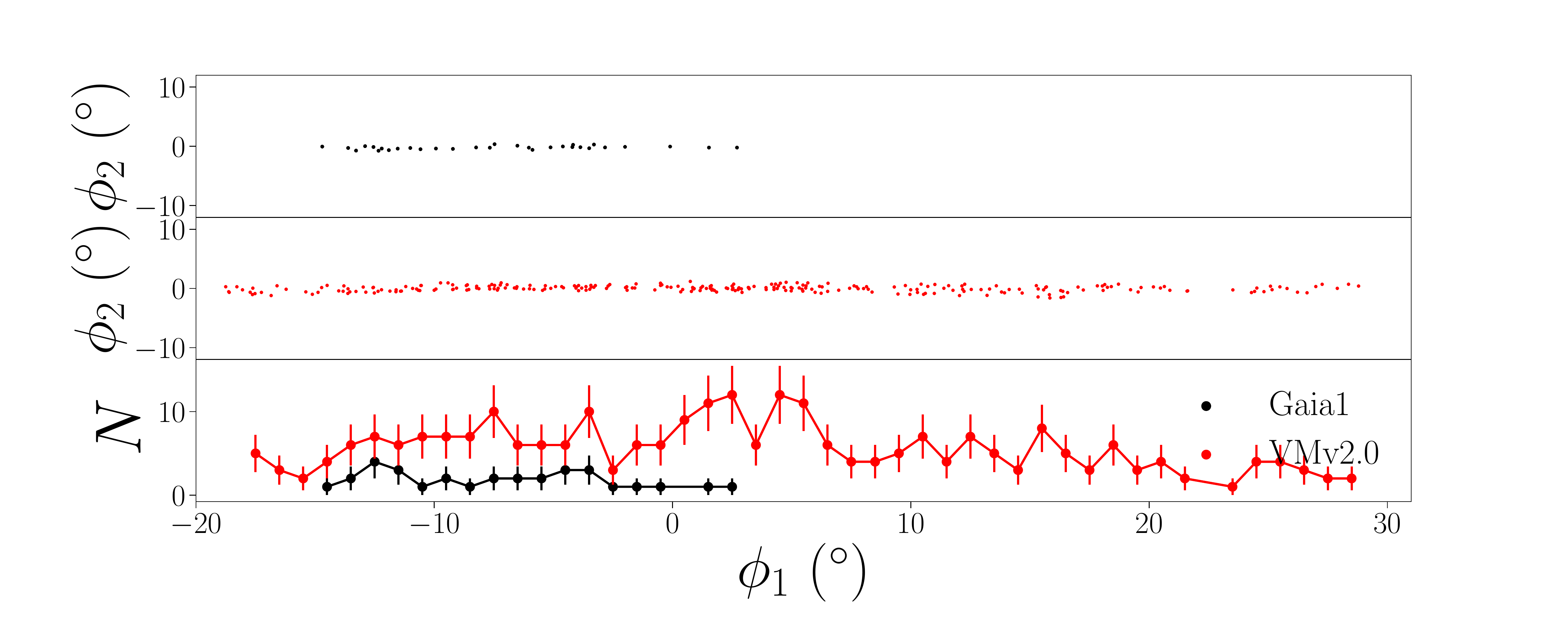}
\includegraphics[width=1\columnwidth]{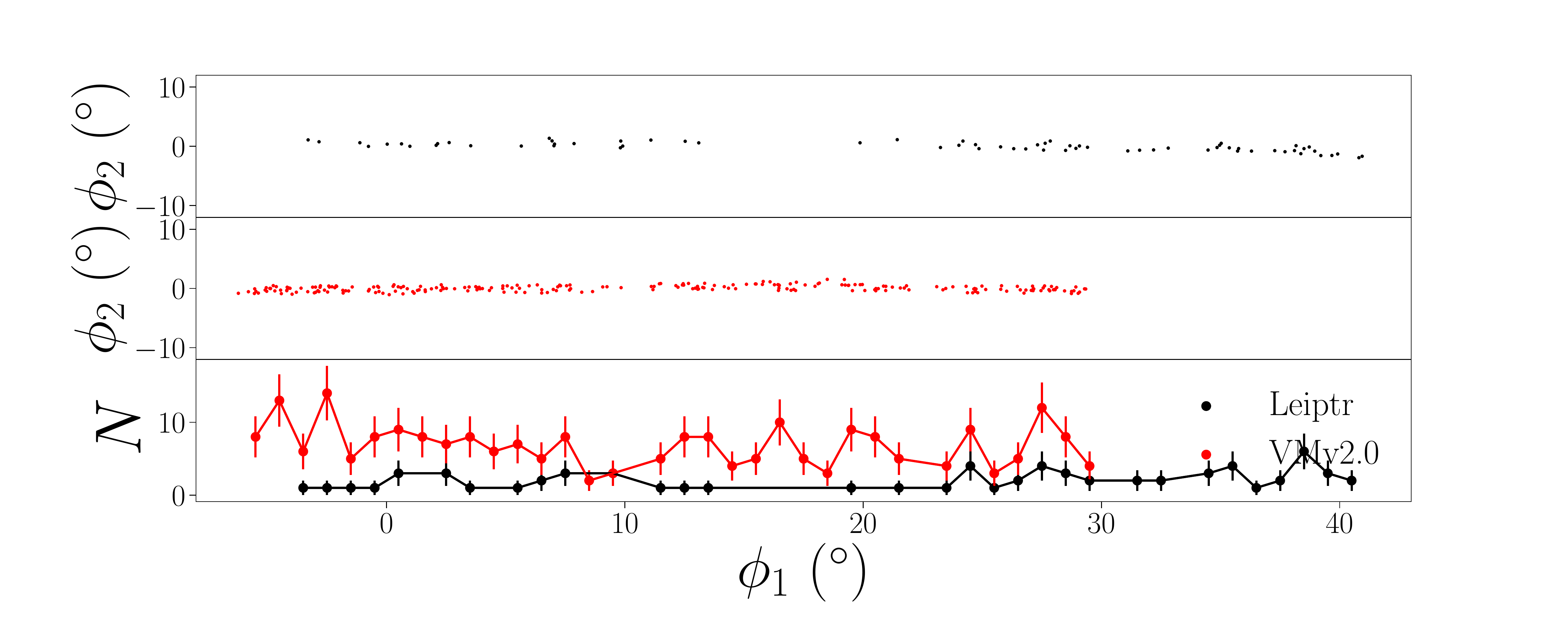}
\includegraphics[width=1\columnwidth]{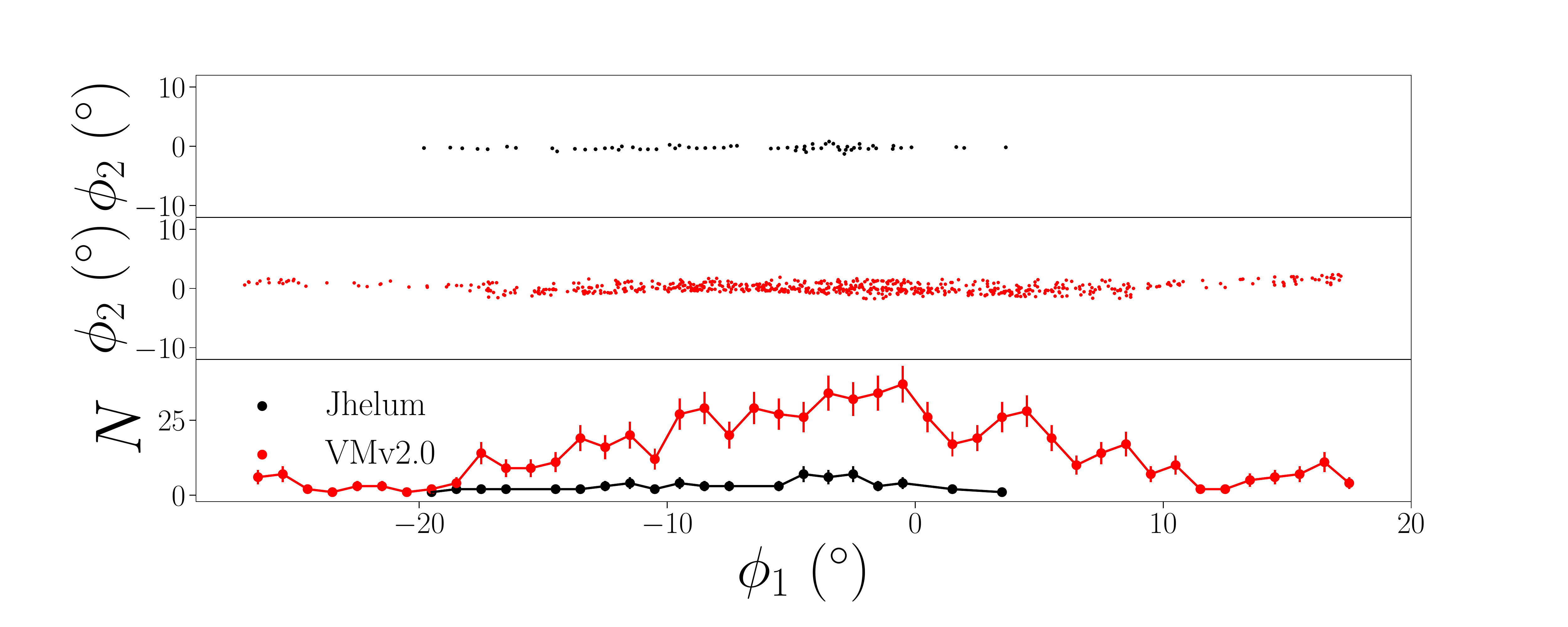}
\includegraphics[width=1\columnwidth]{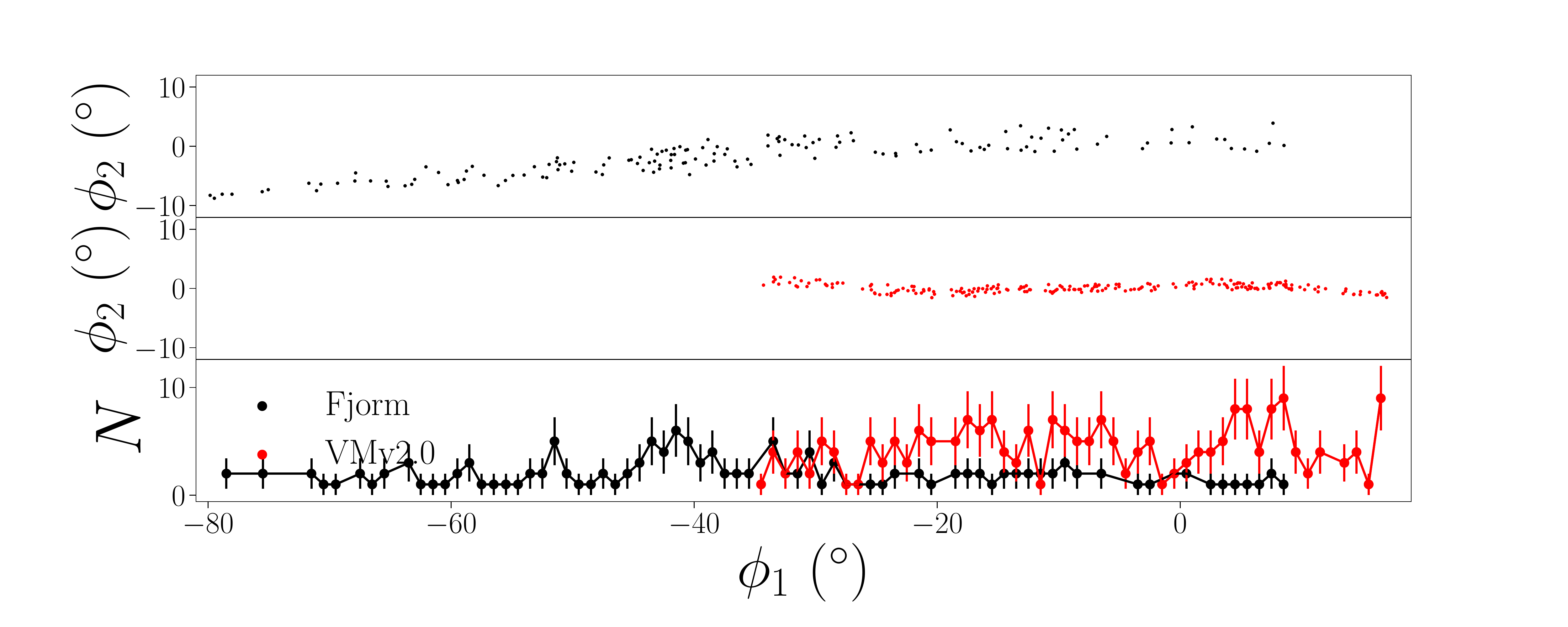}
\includegraphics[width=1\columnwidth]{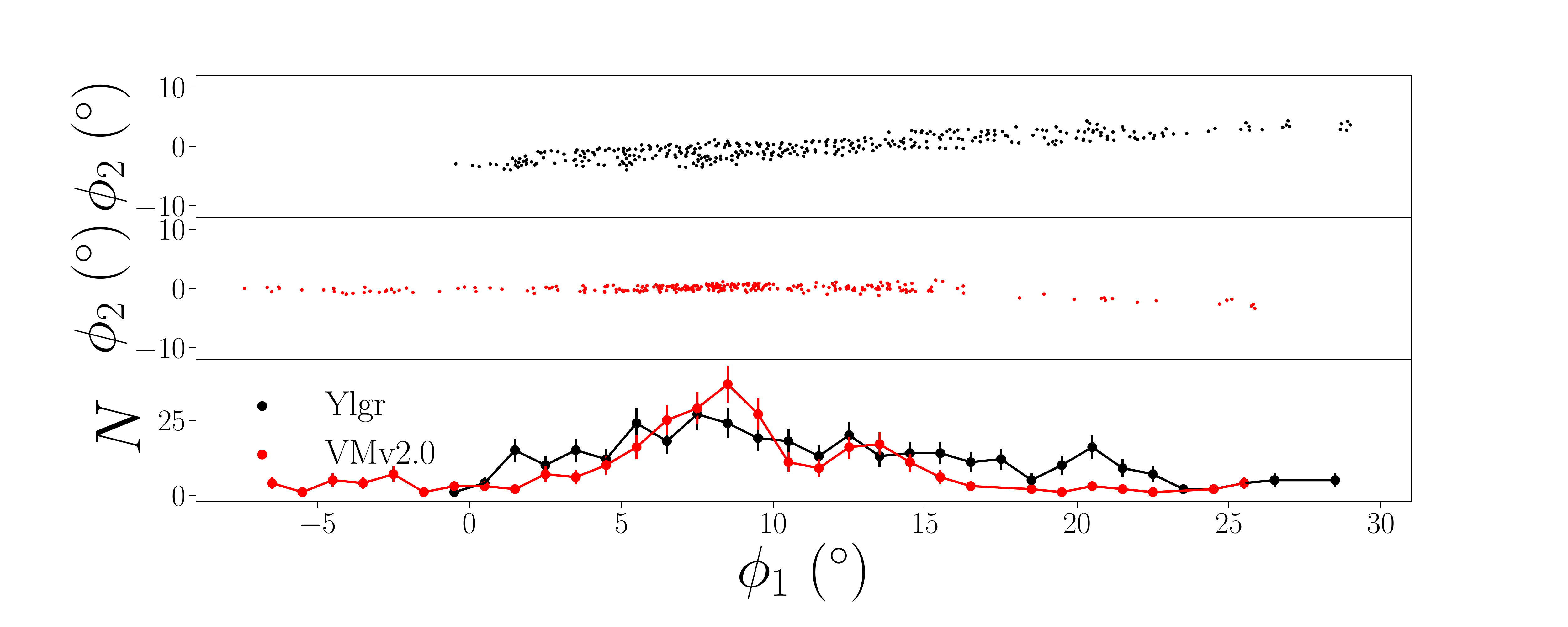}
\caption{Known stellar streams (black, top row) and \vm{}-identified stellar stream candidates (red, middle row) in approximate stream-aligned coordinates $(\phi_1,\phi_2)$ (stream-aligned coordinates are defined separately for each stream). Star counts in $1^\circ$-wide bins of $\phi_1$ are shown in the bottom row. 
}
\label{fig:knownstreams_binned}
\end{centering}
\end{figure*}

A comparison of the positions and proper motions as identified by \streamfinder{} and \via{} is provided in Figure~\ref{fig:streamfinder}. The magnitudes and colors of the stars which \vm{} identifies as members of these known streams are shown in Figure~\ref{fig:known_streams_colormag}. For GD-1, we also show (with a shaded red region) the range of \Gaia{} DR2 stellar colors and magnitudes for the likely stream  members identified by \cite{2018ApJ...863L..20P}. This provides an estimate for the dispersion of of color and magnitudes that might be observed for stellar streams in \Gaia{}. It indicates that \Gaia{} DR2 photometry is insufficiently precise to resolve a clean isochrone in GD-1, and thus we should probably not expect to see one in the other \vm{} stream candidates as well. 

In Figure~\ref{fig:knownstreams_binned}, we plot the binned number of stars as a function of the stream coordinate $\phi_1$,\footnote{The stream-aligned coordinates are obtained by rotating the angular positions of the stars to new coordinate system $(\phi_1,\phi_2)$, which minimizes the sum of $|\phi_2|$ over all the \vm{} stars. The rotation step uses the \textsc{Gala} Python package \citep{gala,gala:zenodo}.} comparing our results to those from \streamfinder{}.
From this figure, we see that some our re-identified streams appear to be significantly extended by \via{} in various directions. However, we note that these extensions are sensitive to the hyperparameters of \via{}, in particular the acceptable angle between protostreams at the combination step. Thus more robust confirmation of these potential stream extension awaits further study. In future work, we will study the orbits of these extensions as well as the chemical abundances of these members to check whether or not these stars are part of the streams. %

From Figure~\ref{fig:knownstreams_binned}, we see that \vm{} does not identify the full extent of Leiptr or Fjorm as found by \streamfinder. For Leiptr we have traced this back to the fact that the missing segment of the stream is fully contained in a patch that was deemed too close to the disk and was cut out of our \vm{} scan (see Section~\ref{sec:dataoverview} for details). For Fjorm, we found that the ANODE step of our algorithm actually correctly identified the remainder of Fjorm as anomalous stars, but the missing parts are too broad for the narrow line-finder settings used in this work (which are applied in the Hough coordinates after the ANODE training is completed). Repeating the analysis for a wider stream setting will be an interesting direction to explore in future work, potentially highlighting a new set of wider stellar streams. 

We also note from Figure~\ref{fig:knownstreams_binned} that \vm{} tends to find a much higher density of stars than \streamfinder{} where the two methods do overlap. The increase in membership stars will enable more spectroscopic fellow-ups, which will improve the characterization of these streams. 

Finally, with Ylgr, there seems to be a mismatch in the direction of the stream between \via{} and \streamfinder --  overall, \streamfinder's Ylgr seems to be at an angle relative to \via's, more clearly seen in Fig.~\ref{fig:streamfinder}. This will be interesting to study further and to check whether the two streams are indeed Ylgr, or they are independent structures that happen to overlap.

Next we turn to the stream candidates found by \streamfinder{} that were not re-identified by \vm{}. For these streams, we define two categories: The first category  -- Indus, Fimbulthul, Sylgr, and Slidr -- have a large number of stars (150, 309, 103, and 156, respectively), while the second category -- consisting of Gaia-4, Gaia-5, and Svol -- have relatively few stars per the \streamfinder{} membership catalog (8, 37, and 45, respectively). %

Investigating the first category, we discover that actually Fimbulthul, Sylgr and Slidr are tagged at the ANODE step, i.e.,\ the 100 highest $R$ stars in their respective ROIs do coincide with the position and proper motions of these \streamfinder{} streams. However, these streams, like the previously-mentioned segment of Fjorm, are again too wide in position space, so they fail to be picked up by the post-ANODE line-finder on the narrow stream setting. This will be explored in more detail in a future work. 

For the second category of missing streams, we find no trace of Gaia-4, Gaia-5 and Svol in any step of \vm, not even after running ANODE. Given the low number of stars within each stream, it is likely these are simply below the detection threshold of ANODE.

Finally, we also find no trace of Indus, despite its robust star count and relative narrowness. It is possible that this failure mode is not attributable to any single property of the stream. However, we note that Indus is composed primarily of dim stars (see Figure~4 of \cite{2018MNRAS.481.3442M}), near the magnitude limit we adopt for \Gaia{} ($g=20.2$). It is possible that the increased number of background stars in this region of phase space is the reason ANODE did not flag the Indus stars as anomalous.

It is notable that the six streams (GD-1, Gaia-1, Leiptr, Jhelum, Fjorm, Gaia-3/Ylgr) identified by both \streamfinder{} and \vm{} are also the highest significance stream candidates according to \vm{}. (Their significances are $\sigma_{\rm stream}=83.0$, $46.9$, $29.2$, $29.1$, $22.6$ and $22.5$, respectively.) Given the differences in the methodologies of \streamfinder{} and \via{} and their agreement on the highest significance streams, we conjecture that these six streams are the most robust thin cold streams in the volume of \Gaia{} data analyzed in this work.

\subsection{New stream candidates}
\label{sec:newstreams}

Finally, we turn to stellar stream candidates that have not been previously identified (referencing against the \textsc{Galstreams} database \citep{2022arXiv220410326M}).
There are a total of 102 \vm{} stream candidates by \vm{} with significance $\sigma_{\rm stream}\geq 8.86$, the cutoff we adopt based on our comparison with \Galaxia{} (see Sec~\ref{sec:galaxia}). 
This number includes GD-1, the five other previously identified streams, and the four possible Sagittarius stream components.

\begin{figure}
\begin{centering}
\includegraphics[width=0.85\columnwidth]{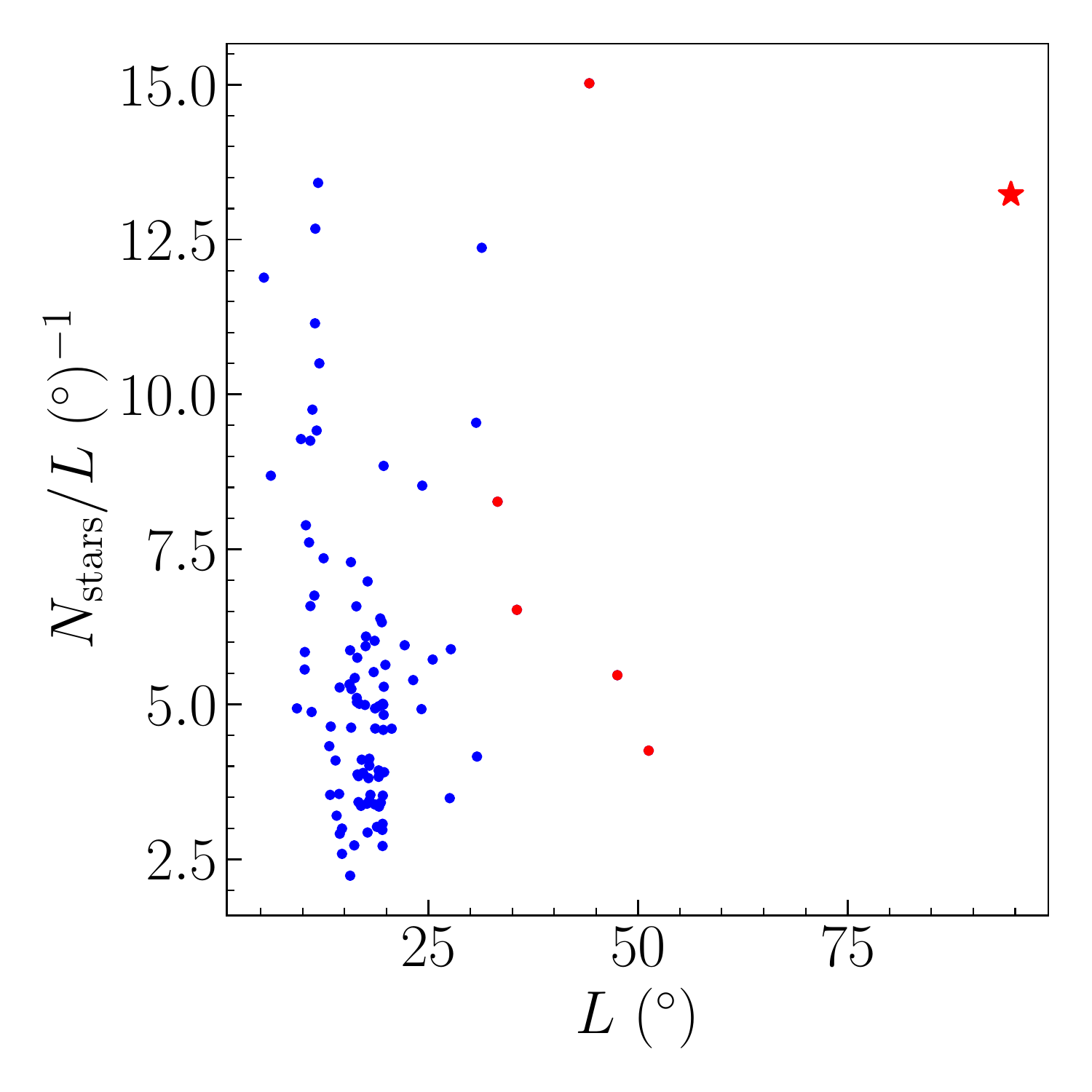}
\caption{Length $L$ of stellar stream candidates as identified by \vm{} versus number of \via{}-identified member stars per unit length $N_{\rm stars}/L$ for all 102 \vm{} stream candidates. Stream candidates that correspond to streams previously identified by \streamfinder{} are shown in red. GD-1 is denoted with a red star.
\label{fig:stream_length_density}}
\end{centering}
\end{figure}

As to why the majority of our stream candidates have not been previously discovered by \streamfinder{}, we can offer a few possible explanations: 
\begin{itemize}

\item One possibility is the fpr is being  underestimated due to some systematic mis-modeling by the \Galaxia{}. We deem this unlikely, given how well the \Galaxia{} seems to be reproducing the bulk properties of the ROIs and protoclusters found by \Gaia{} DR2. Nevertheless, it would be important to perform more cross-checks and estimates of the fpr. For some ideas on how to do this in the future, see Section~\ref{sec:conclusions}.

\item A second possibility is that these stream candidates tend to be shorter on average than the typical \streamfinder{} stream, which may impact their detectability using the \streamfinder{} algorithm. This is indicated in Figure~\ref{fig:stream_length_density}, where we plot the density of the stream candidates as a function of their length, and single out the streams found by \streamfinder{} in red.  

\item Also, \streamfinder{} only reported on a subset of their stream candidates. It is possible that our new stream candidates were also found by \streamfinder{}, but just were not made public.

\item A final possible explanation is that the Galactic potential modeling assumptions of \streamfinder{} are being violated for these streams. We expect this behavior to be affecting mostly the southern hemisphere streams, due to the proximity to the Large Magellanic Cloud LMC (see e.g. \cite{2021ApJ...923..149S,2022arXiv221104495K,2023MNRAS.518..774L} for studies on the effect of the LMC on the Milky Way potential). To this point, we note that \streamfinder{} has a clear bias towards the Northern Galactic hemisphere, which could be related to the influence of the LMC/SMC on the Galactic potential model, whereas \vm{} is fairly symmetric between Northern and Southern hemispheres.
\end{itemize}

\begin{figure*}
\begin{centering}
\includegraphics[width=2\columnwidth]{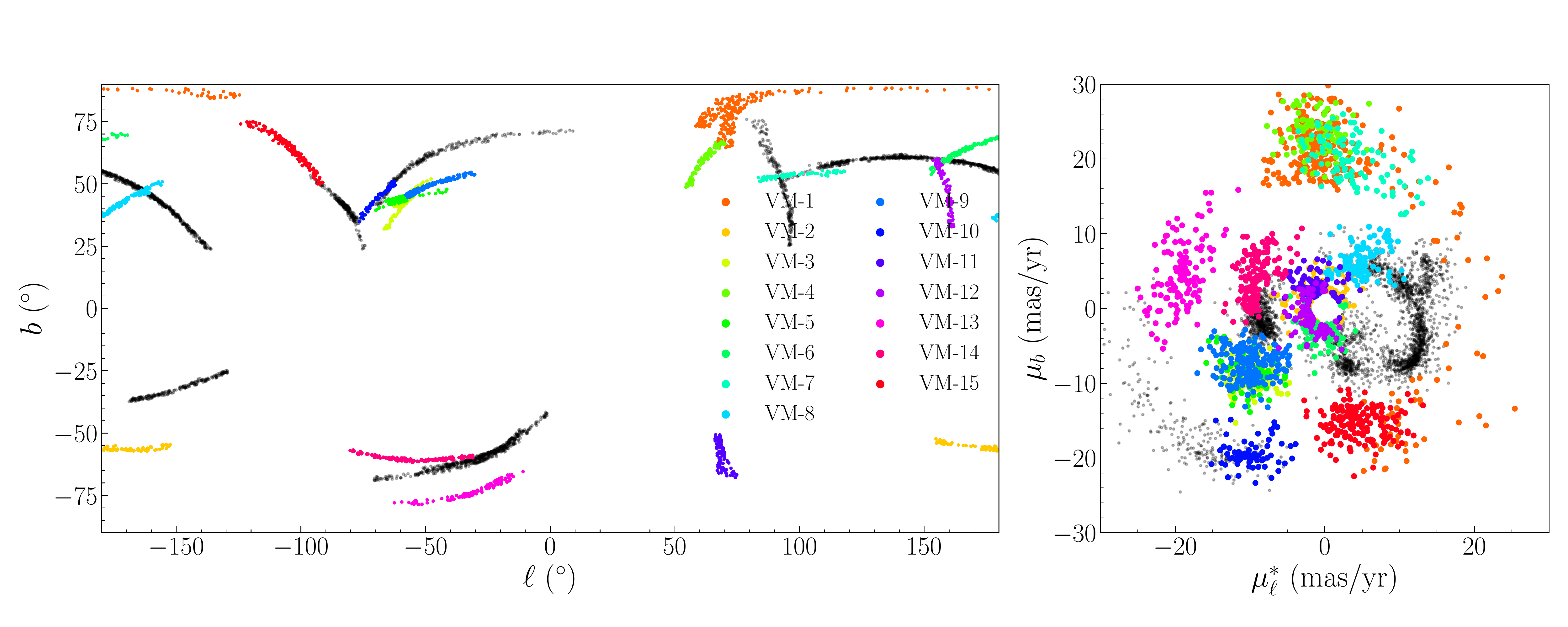}
\caption{The 15 stream candidates identified by \vm{} which are composed of protoclusters from more than one patch in Galactic position (left) and proper motion (right). Stars belonging to the new stream candidates are identified by colored points; previously-identified streams which have been re-identified in \vm{} are shown in black (see Sections~\ref{sec:gd1} and \ref{sec:rediscovery}, and Figure~\ref{fig:streamfinder}). 
\label{fig:new_streams}}
\end{centering}
\end{figure*}

\begin{figure*}
\begin{centering}
\includegraphics[width=2\columnwidth]{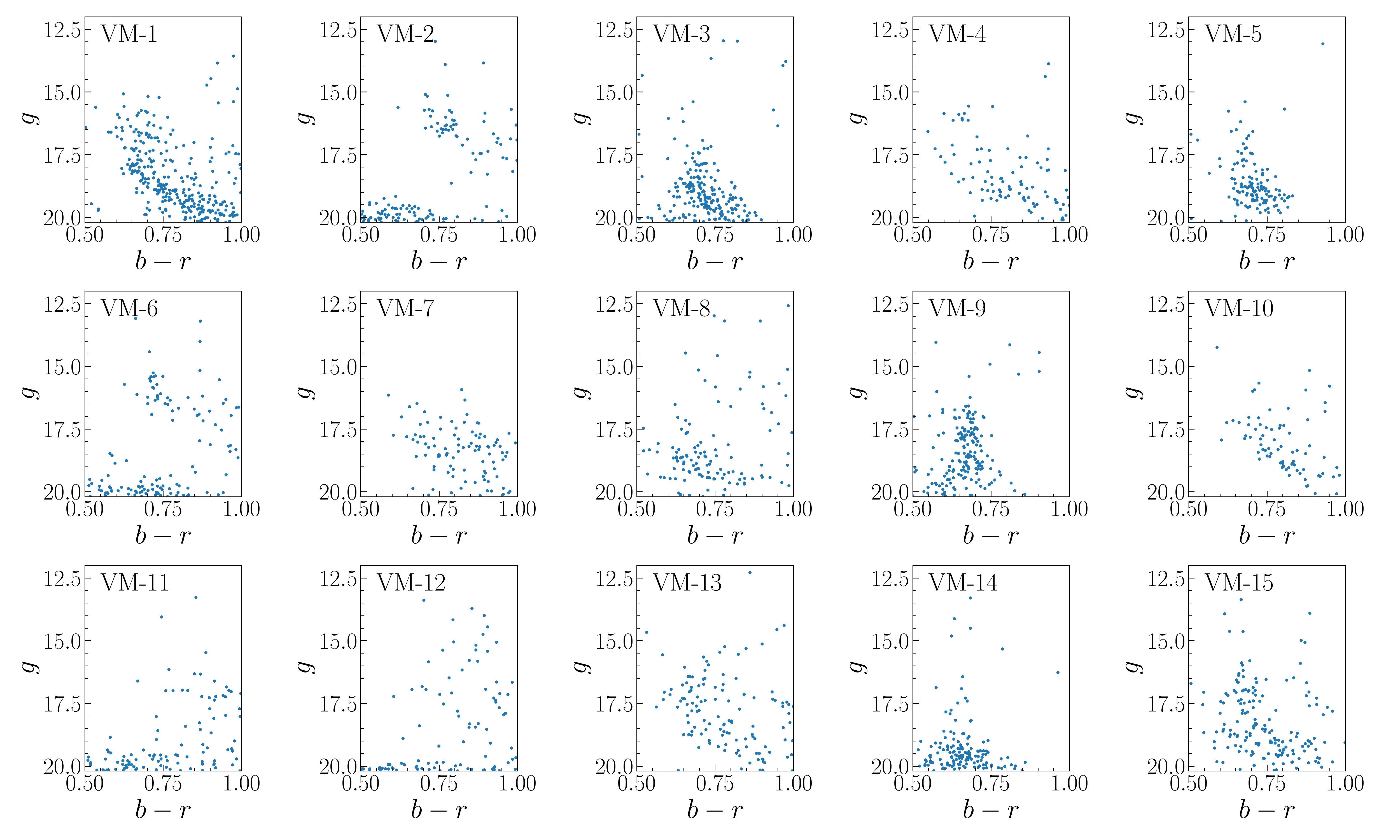}
\caption{The 15 highest-significance stream candidates identified by \vm{} which are not identified with previously known streams, in color $b-r$ versus magnitude $g$.
\label{fig:new_streams_colormag}}
\end{centering}
\end{figure*}

In Figure~\ref{fig:new_streams}, we show the locations and proper motions of the 15 highest-significance \via{} new stream candidates that do not overlap with Sagittarius (only one possible Sagittarius stream component also has a significance as high as these 15). These top candidates have $\sigma_{\rm stream}$ values between 11.6 and 19.9, and include all candidates composed of protoclusters drawn from more than one patch on the sky. Only two of these 15 candidates are composed of stars from a single patch (the $9^{\rm th}$ and $10^{\rm th}$ most significant stream candidates). The photometric properties of the constituent stars are shown in Figure~\ref{fig:new_streams_colormag}. Stream candidate indexing goes in decreasing stream significance. We chose to highlight these candidates as -- due to their length, high $\sigma_{\rm stream}$ values, and presence in multiple patches -- we judge them to be among the most robust candidates in our sample. These, as well as the other stream candidates, require additional cross-matched observations to confirm or reject their existence as stellar streams.

\begin{figure}
\begin{centering}
\includegraphics[width=0.9\columnwidth]{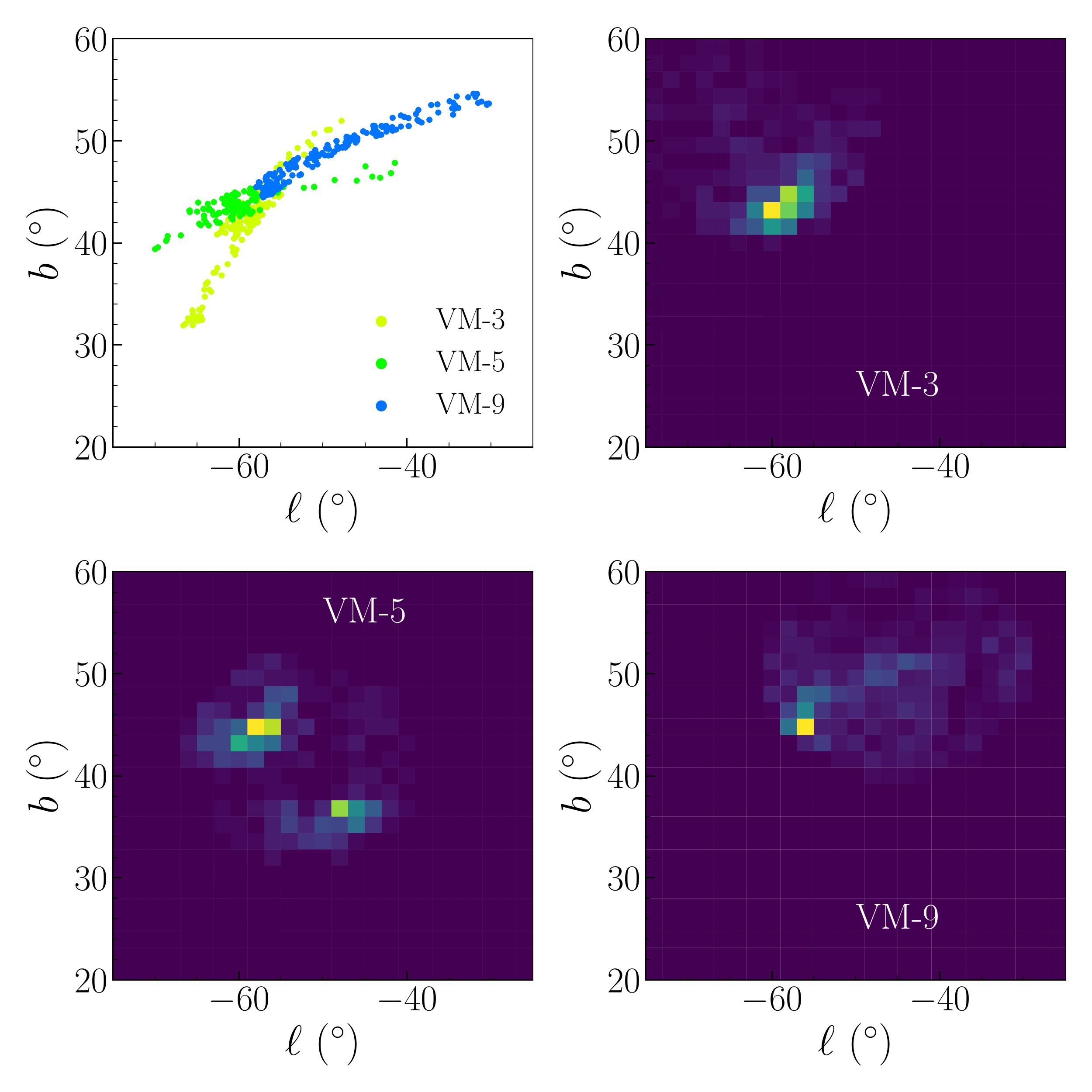}
\caption{Upper left: the stars in the highest-significance protoclusters of the three stream candidates (VM-3, 5, and 9) which cross in angular position. Clockwise from upper left: two-dimensional histograms of all the stars in all the protoclusters in each stream, clearly showing the shared clump of stars around which each stream candidate is built.
\label{fig:stream_clump}}
\end{centering}
\end{figure}

One can see in Figure~\ref{fig:new_streams} that three of the 15 highest-significance stream candidates (VM-3, 5, and 9) are overlapping in angular position and proper motion. Upon closer examination (see Fig.~\ref{fig:stream_clump}), we discover that all three streams appear to be built around the same stellar overdensity. Though a stream-like extended structure can be seen extending from this object (most clearly in VM-9), differences in the highest-$R$ stars among each of the streams resulted in three different stream candidates with different paths across the sky. As a result of these path differences, these three stream candidates were not merged together. Further investigation of the stars around which these three stream candidates are built is necessary; it does not seem to correspond to a known object.

\section{Conclusions }
\label{sec:conclusions}

In this work, we have described \vm{}, an update to the model-agnostic, fully-automated stream finding algorithm \via. \via{} is based on the weakly-supervised anomaly detection protocol ANODE \citep{Nachman:2020lpy}, originally developed for resonant anomaly detection at the Large Hadron Collider (LHC). \via{} takes as inputs only the sky positions, proper motions, colors, and magnitudes of stars in the \Gaia{} catalog, and searches for stream-like overdensities in this space in a model-agnostic way. 

Our work builds upon a previous version of \via{} described in  \cite{2022MNRAS.509.5992S}, which focused on a more limited demonstration of finding the well-known stream GD-1 in a model-agnostic way.  Here, we generalize the application of \via{} to the entire \Gaia{} DR2 data set. This required a number of improvements to the algorithm. The core step remained the same: training ANODE on {\it signal regions} and {\it sideband regions} consisting of windows in a single proper motion coordinate and their complements, in order to derive an ``overdensity score" or ``anomaly score" for each star in the signal region. However, the subsequent steps of combining multiple detections of stream fragments into a coherent and consistent high-significance stream candidate were redesigned in order to make the algorithm more robust to the different streams. We have also improved on the version of \via{} presented in \cite{2022MNRAS.509.5992S} by incorporating a second ANODE scan over the orthogonal proper motion coordinate, and improving the data quality with a more sensitive globular cluster / dwarf galaxy finding algorithm and a new fiducial cut that excludes foreground contamination from bright stars originating in the thick disk of the Milky Way.

We have validated \via{} on a combination of \Gaia{} DR2 data and the \Galaxia{} simulation, an idealized simulation of the Milky Way that, by design, does not include any substructure.  In particular, using \Galaxia{}, we estimated the fpr of stream detections in \via. After showing that this fpr is at an acceptable level ($\sim 10\%$ for 102 detected stream candidates), we applied \via{} to a full-sky scan of \Gaia{} DR2 data. We showed how \via{} could rediscover the well-known GD-1 stream, as well as five other streams previously detected in \Gaia{} DR2 by the \streamfinder\ method (Gaia-1, Gaia-3/Ylgr, Fjorm, Leiptr and Jhelum). Finally, we described and characterized the 90+ additional stream candidates that \via{} finds, focusing in particular on 15 of the highest-significance new stream candidates. 

Although this current paper focused on \Gaia{} DR2, a major priority is to re-run it on the \Gaia{} Data Release 3 (DR3) data. We expect, with the improved measurements of DR3, all the stream detections will become even more significant and robust, and perhaps discoveries of even more new stream candidates await.

Our work presents many directions for further study. Foremost among these are confirming our new stream candidates with in-depth studies of orbital parameters as well as their chemical abundances, which will help identifying their origins \citep[similar to what][has done for some of the \streamfinder{} streams]{2022MNRAS.516.5331M}. If some of these structures are remnants of dwarf galaxy mergers, it will lead to a better understanding of the Milky Way's stream population \citep{2022ApJ...928...30L}, which is crucial for understanding the merger history of the Milky Way, the rates of disruptions, and the feedback models in numerical simulations \citep[as illustrated for example in the comparative work of][]{2022arXiv220802255S}.    

Besides re-running \vm{} on \Gaia{} DR3 and performing more in-depth astrophysical studies of our stream candidates, it is important to perform additional checks on the stream candidates and the fpr estimated in this work.
While re-running ANODE on the entire DR2 sky multiple times is computationally unfeasible, it should be possible to re-run on just those patches or SRs that contain the new stream candidates. If the stream candidate is found with repeated ANODE scans, this makes it more likely to be real. Finally, repeating the fpr study with a different simulation would provide a very important cross check. In the future, \Gaia{} mock catalogs based on state-of-the-art, kinematically consistent hydrodynamical simulations could become viable candidates for such a study \citep{2022arXiv221111765L}.

Our findings also suggest that several of the previously discovered streams may be longer than previously known. We will address the extended streams in upcoming work by comparing the chemical abundances of the newly found members with the previously confirmed members. These extensions will be helpful in better studying the orbital properties of these streams, which will then lead to updated measurements of the properties of the Milky Way halo, such as its total mass and shape, as well as the effect of the LMC.

Looking further beyond the present work, potentially one of the most interesting findings are the many stream candidates discovered by our algorithm and not reported by \streamfinder{}. 
If these stream candidates are indeed real and not false positives (which we expect given our estimated low fpr presented in this paper), this could indicate that the model-dependent methodology of \streamfinder{} is causing it to miss many streams.  
This would be the ultimate benefit of a model-agnostic technique such as \via.

Beyond this work, a few generalizations of \via{} will reveal more of the interesting merger history of Milky Way. In particular, we want to highlight the wide stream search, where by inspection, some of the streams from the literature are being missed in our analysis due to the settings on the width of the stream. 
Improving the machine-learned anomaly detection step currently performed by ANODE may also reduce the false positive rate and/or be more sensitive to other aspects of the stellar streams. For example, the CATHODE anomaly detection algorithm \citep{Hallin:2021wme} has been shown to improve anomaly detection within LHC data, and may do the same here. More light-weight anomaly detection methods, such as CWoLa-Hunting~\citep{Collins:2018epr,Collins:2019jip}, could be used to improve refine and improve stream membership post-discovery. 
Finally, this method will be of upmost importance to future surveys, in particular the Vera Rubin Observatory Legacy Survey of Space and Time (LSST) \citep{2009arXiv0912.0201L,2019ApJ...873..111I}, which will not only improve measurements of proper motions, but will unveil a deeper side of the sky.

\section*{Acknowledgements}
\addcontentsline{toc}{section}{Acknowledgements}

We would like to acknowledge to John Tamanas for collaboration in the early stages of this work.
We would also like to thank Anna Hallin, Claudius Krause, Ben Nachman, Mariel Pettee, Nora Shipp and Sowmya Thanvantri for helpful discussions. Finally, we are grateful to Anna Hallin and Claudius Krause for feedback on the draft.
MB and DS are supported by the DOE under Award Number DOE-SC0010008.
LN is grateful for the generous support and hospitality of the Rutgers NHETC Visitor Program, where this work was initiated.

This research used resources of the National Energy Research Scientific Computing Center (NERSC), a U.S. Department of Energy Office of Science User Facility operated under Contract No. DE-AC02-05CH11231.

This work has made use of data from the European Space Agency (ESA) mission
\Gaia{} (\url{https://www.cosmos.esa.int/gaia}), 
processed by the \Gaia{}
Data Processing and Analysis Consortium (DPAC,
\url{https://www.cosmos.esa.int/web/gaia/dpac/consortium}). 
Funding for the DPAC
has been provided by national institutions, in particular the institutions
participating in the \Gaia{} Multilateral Agreement.


\section*{Data Availability}

This paper made use of the publicly available \Gaia{} DR2 data. For more detailed information about the data (e.g.~patch center locations) please contact the corresponding author. Membership stars of the discovered streams will be provided following journal acceptance.

\appendix

\section{Binned Two-Dimensional Overdensity Detection}\label{app:annulus}

In this work, we find it necessary in two contexts to identify local overdensities in two-dimensional coordinates: first when locating GC/DGs in angular coordinate-space, and second in finding line-parameters in the $(\rho,\theta)$ Hough space. In both cases, the problem at hand is locating the coordinates of the highest-density of points in a two-dimensional space, over a background with is not at all uniform and cannot be predicted {\it a priori}. We use a single flexible algorithm for both situations, modifying the hyperparameters to address each of these two cases. 

This algorithm operates on a 2d histogram consisting of a set of pixels indexed by $i$ and $j$, each with a number of counts $N_{ij}$. We are interested in whether one of the pixels is elevated (overdense) relative to its neighbors. The main ``hyperparameter" of this algorithm is how the neighbors are chosen in order to estimate the background level. Given some list $(i_1,j_1)$, $(i_2,j_2)$, $\dots$ of neighbors of the $(i,j)$ pixel, let ${\cal A}_{ij}$ be their counts:
\begin{equation}
    {\cal A}_{ij} = \{ N_{i_1j_1},\ N_{i_2j_2},\,\ldots\}.
\end{equation}
The position-dependent background estimate is then the average of the counts in this neighbor list:
\begin{equation}
    N_{{\rm bg},ij} = \bar{\cal A}_{ij}.
\end{equation}
We also define a ``systematic error'' in our background estimate in terms of the standard deviation of the neighbor counts $\sigma({\cal A}_{ij})$ and the number of pixels in the neighbor list $|{\cal A}_{ij}|$:
\begin{equation}
    \delta N_{{\rm bg},ij} = \sigma({\cal A}_{ij})/\sqrt{|{\cal A}_{ij}|}.
\end{equation}
This allows us to define a significance for each pixel $i,j$:
\begin{equation}
    \sigma_{ij} = \frac{N_{ij}-N_{{\rm bg},ij}}{\sqrt{ N_{{\rm bg},ij}+(\delta N_{{\rm bg},ij})^2}}.
\end{equation}
where the Poisson statistical error and the systematic error are added in quadrature.

For the GC-finding algorithm, the neighbor list is an $11\times 11$ pixel annulus centered at $i,j$ with the central $3\times 3$ pixels removed. 

For the line-finding algorithm, the neighbor list is more complicated. Here each pixel $(i,j)$ counts the number of curves in Hough space passing through a box of width $\Delta i=5$ and $\Delta j=3$, centered at $\rho=-10+i/5$ and $\theta=\pi\times j/100$. Thus adjacent pixels correspond to overlapping boxes and are not independent of one another. To derive the background counts with independent boxes, we use a $7\times 7$ annulus of pixels centered at $(i,j)$, spaced apart by $\Delta i$ and $\Delta j$, with the middle $3\times 3$ pixels removed. 

\begin{figure}
\includegraphics[width=.95\columnwidth]{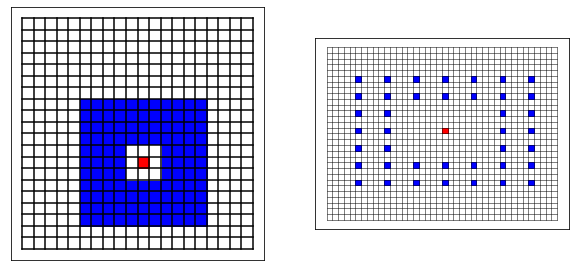}
\caption{Left: the $11\times 11$ annulus used for estimating the background level in the GC-finding algorithm. Right: the $7\times 7$ annulus with stride $5\times 3$ used for estimating the background level in the line-finding algorithm.}
\label{fig:gridpixels}
\end{figure}

These neighbor lists for the GC-finding and the line-finding are shown in Figure~\ref{fig:gridpixels}.

\bibliographystyle{mnras}
\bibliography{streamfinding} 



\bsp	
\label{lastpage}
\end{document}